\documentclass[logo,url,11pt,a4paper]{ETHpaper}
\pdfoutput=1
\usepackage{amssymb,amsmath} 
\usepackage{subfigure}
\usepackage[sort&compress]{natbib} 
\usepackage[utf8]{inputenc}

\begin{document}

\title{A framework for cascade size calculations on random networks}
\author{Rebekka Burkholz, Frank Schweitzer}
\authoralternative{R. Burkholz, F. Schweitzer}
\address{Chair of Systems Design, ETH Zurich, Weinbergstrasse 58, 8092
   Zurich, Switzerland}

\reference{Version of \today }

\www{\url{http://www.sg.ethz.ch}}

\makeframing
\maketitle

\begin{abstract}
We present a framework to calculate the cascade size evolution for a large class of cascade models on random network ensembles in the limit of infinite network size. 
Our method is exact and applies to network ensembles with almost arbitrary degree distribution, degree-degree correlations and, in case of threshold models, for arbitrary threshold distribution.
%
%
%
With our approach, we shift the perspective from the known branching process approximations to the iterative update of suitable probability distributions.
Such distributions are key to capture cascade dynamics that involve possibly continuous quantities and that depend on the cascade history, e.g. if load is accumulated over time. 
These distributions describe respect the Markovian nature of the studied random processes. 
Random variables capture the impact of nodes that have failed at any point in the past on their neighborhood. 
As a proof of concept, we provide two examples: (a) Constant load models that cover many of the analytically tractable casacade models, and, as a highlight, (b) a fiber bundle model that was not tractable by branching process approximations before.
Our derivations cover the whole cascade dynamics, not only their steady state.
This allows to include interventions in time or further model complexity in the analysis.

\end{abstract}

\section{Introduction}

\subsection{Network representation}

Many models in Statistical Physics describe the dynamics of interacting particles that are characterized by a binary state, e.g. $\{0,1\}$, healthy/failed, etc.   
Examples include the zero-temperature random-field Ising model \citep{IsingCascade}, percolation models \citep{percolation}, such as the $k$-core percolation model on random graphs \citep{dorogovtsev2006k}, or threshold models \citep{Watts2002,Lorenz2009,Burkholz2015}. 
The Bak-Tang-Wiesenfeld sandpile model \citep{Bak1987} has even become a paradigm for the study of self-organized criticality.

The dynamics of these models can be also interpreted as a cascade process where the change in the state of one particle subsequently impacts the state of the neighbors.  
Such cascade models find applications in different fields, for instance, in systemic risk analyses of financial or economic systems \citep{Schweitzer2009c,Gai2010,BurkholzMultiplex}, information propagation in form of voter models or models for meme popularity \citep{PhysRevX.6.021019}, or simple epidemic spreading processes \citep{RevModPhys.87.925}.

We can represent these systems of interacting particles as a network  (or graph) $G = (V,E)$ consisting of nodes $V$ and links $E$.
Each particle is represented by a node (or vertex), its interactions by links (or edges).
Each node $i \in V$ is characterized by a binary (or in general discrete) state $s_i \in \{0,1\}$ that indicates, for instance, whether $i$ is failed, infected, activated, has a positive spin or adopted a certain opinion ($s_i = 1$). 
A switch of this state involves an action of the node that influences its network neighbors, i.e. the nodes it is connected with. 
This can trigger further state switches and thus lead to a cascade of successive state changes. 
%
%
%
%

Many theoretical investigations of cascade phenomena are concerned with the question how the network topology and 
specific node attributes contribute to an amplification of the dynamics such that large cascades result.
Usually, the cascade size is measured by the fraction $\rho$ of nodes in one state as average over random graph ensembles with given degree distribution. 
In most of the mentioned examples, $\rho$ can be iteratively calculated in the (thermodynamic) limit of infinitely large network size with the help of a branching process approximation, also known as local tree or heterogeneous mean field approximation \citep{RevModPhys.80.1275,Gleeson2007,Burkholz2015}. 
In comparison with Monte Carlo simulations, these calculations usually save computational time and efforts, while they further deepen the theoretical understanding of the key factors driving a cascade.
Commonly, they rely on updating compositions of generating functions which correspond to discrete probability distributions, for instance, the degree distribution of a network.

However, in models where nodes interact via a heterogeneous load redistribution mechanism, this approach breaks down, especially when continuous quantities of load can be distributed that depend on the history of the cascade. 
This is the case, for instance, in fiber bundle models \citep{RevModPhys.82.499}, which have become one of the most prevalent models to describe fracture in materials \citep{newFiber}.
Also neural network models \citep{brette2006exact} and multilayer perceptrons or deep learning architectures \citep{Goodfellow-et-al-2016} involve similar load distribution mechanisms, usually on directed networks. 
To enable their analysis in the thermodynamic limit of infinite network size, we present an alternative view on branching process approximations and shift the perspective towards the iterative update of suitable probability distributions.
Within this analytic framework, we can correctly compute the whole time evolution of the average cascade size for a large class of cascade processes. 
We demonstrate this with the example of a fiber bundle model \citep{moreno2002instability,Lorenz2009} which could, to the best of our knowledge,  not be tackled analytically before. 
Furthermore, we show how our framework simplifies the known local tree approximations in case of threshold models on weighted networks \citep{Burkholz2015}.

While our approach works well for models with local load distributions, we cannot tackle models that require local information, for instance load distribution along shortest paths \citep{Motter2002}, or introduce network clustering \citep{LRD}. 
Yet, our approach is not limited to binary state dynamics.
It applies to the calculation of average node states that can correspond to general discrete or continuous variables.
Yet, we mainly discuss binary state models, as they have motivated our derivations. 

\subsection{Cascade models}

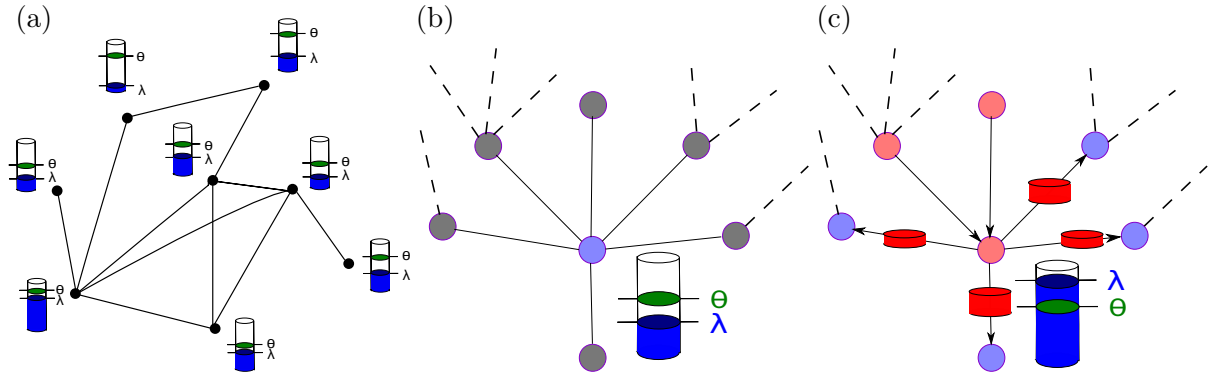
\begin{figure}[t]
 \centering
 \includegraphics[width=0.33\textwidth]{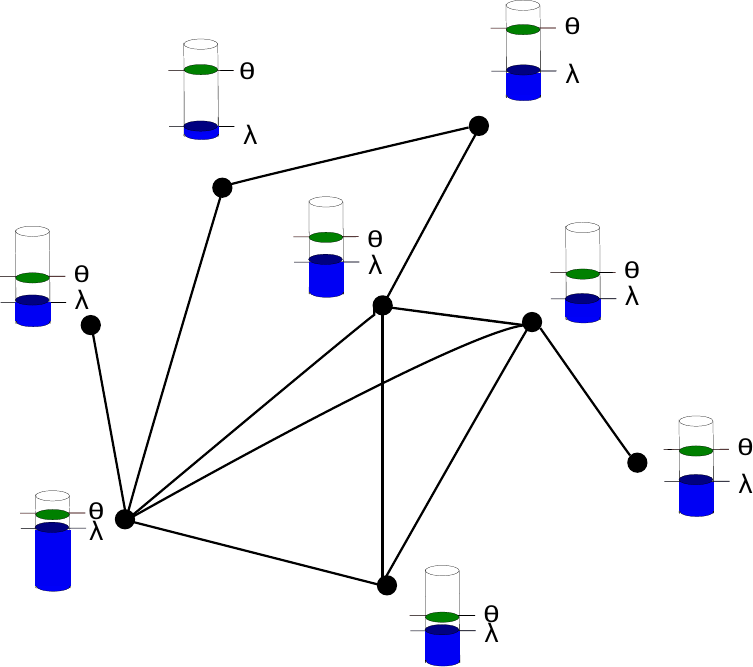}
   \includegraphics[width=0.32\textwidth]{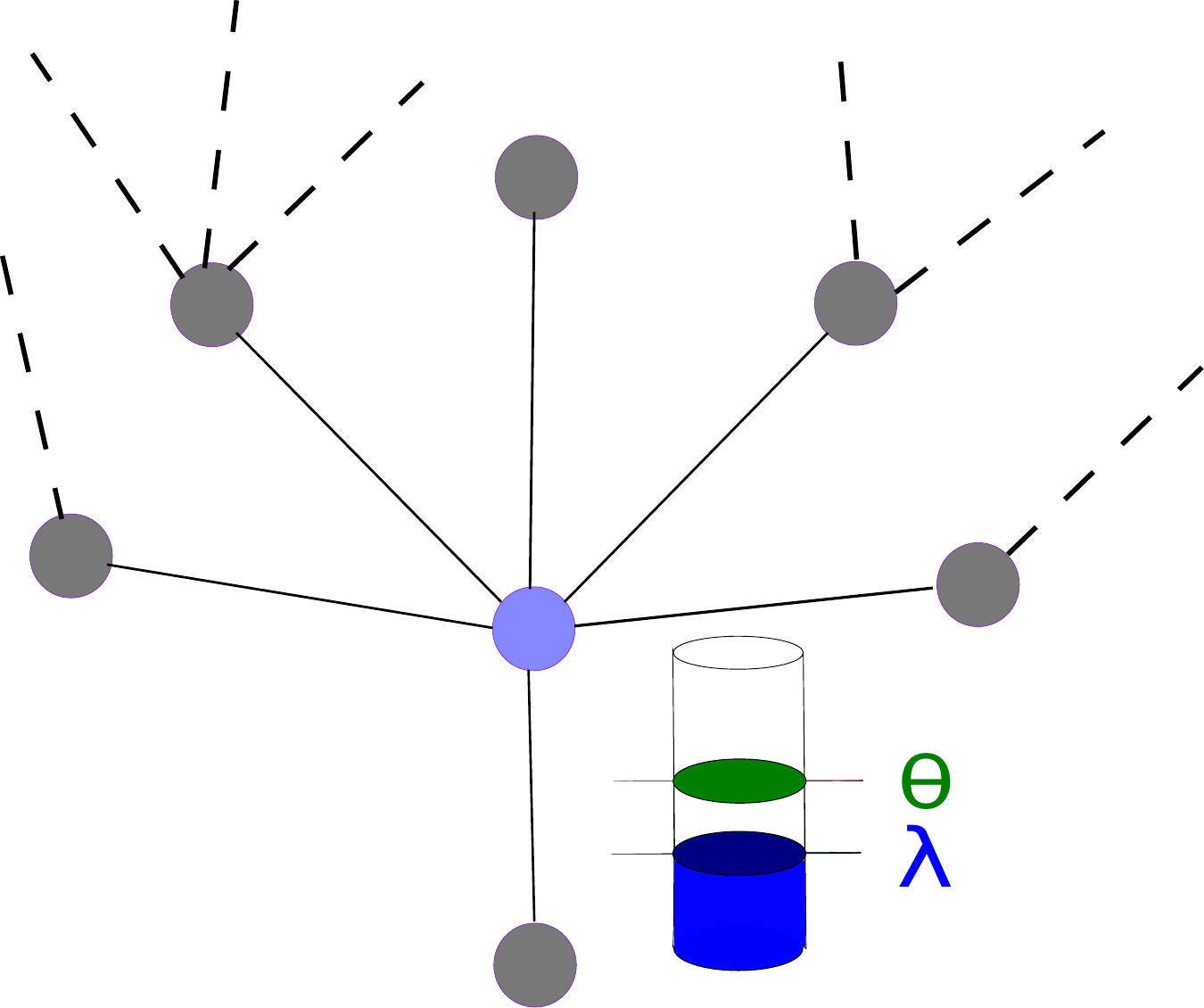}
    \includegraphics[width=0.32\textwidth]{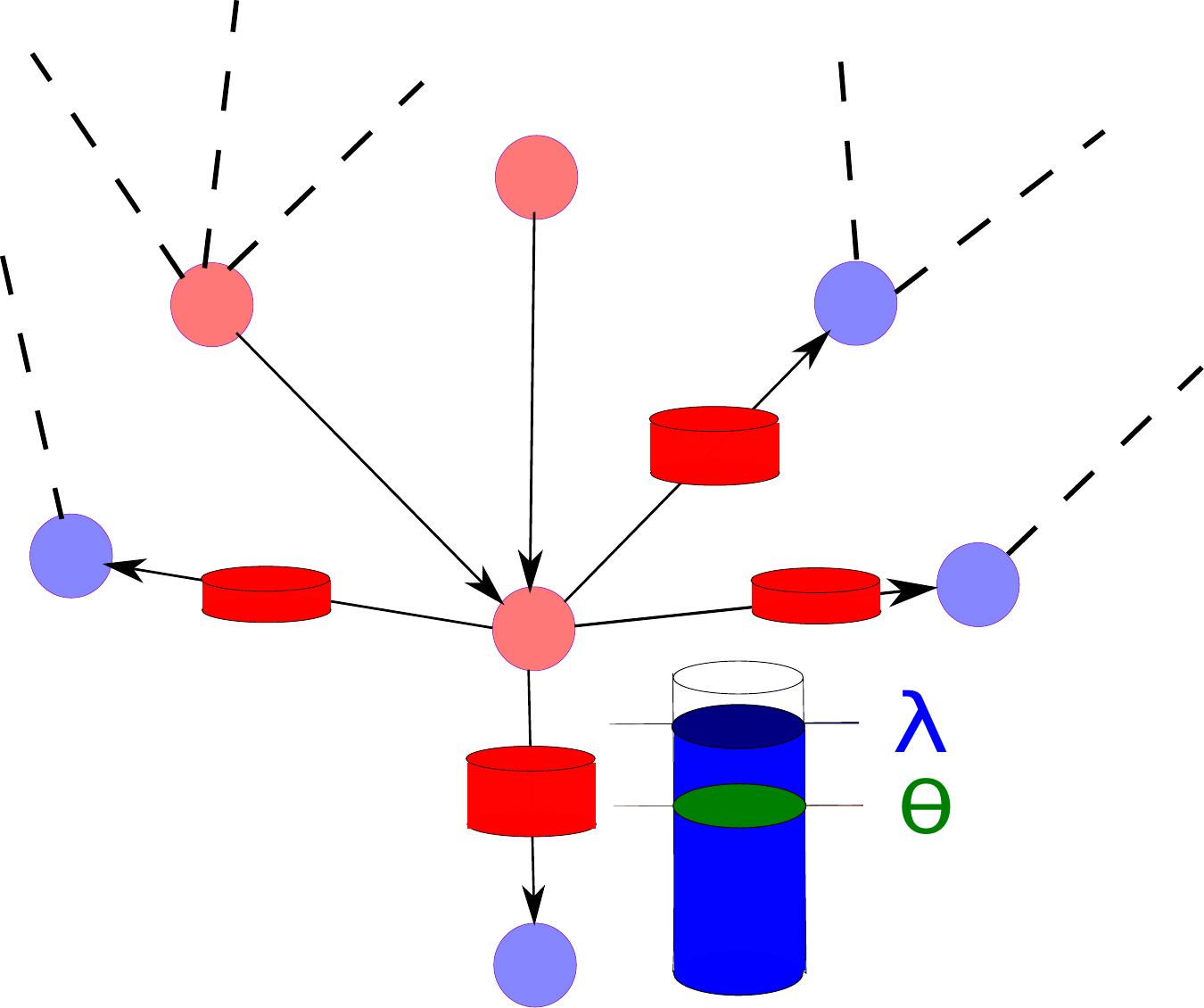}
    \begin{picture}(0,0)
    \put(-450,130){(a)}
    \put(-300,130){(b)}
    \put(-150,130){(c)}
   \end{picture}
 \caption{Illustration of the internal node dynamics. (a) All nodes in the network carry a certain amount of load $\lambda$, which is visualized by a blue liquid in a glass. Additionally, they are equipped with a threshold $\theta$, which is represented by a green disk. Both variables can vary between the nodes and determine a node's state $s$. (b) A node, here colored in blue, is functional ($s = 0$) when $\lambda < \theta$, while (c) it fails ($s = 1$) when $\lambda \geq \theta$. It then distributes load, which is illustrated by red barrels, to its functional network neighbors. A failed node is colored in red, functional ones in blue.}
\label{fig:illustrationFailure}
\end{figure}

The essence of our derivations is in the analytical description of the dynamics for a class of fiber bundle models. 
Therefore, when introducing general cascade processes, we borrow the  wording and the interpretation from the fiber bundle model.
Each node $i \in V$ in a network $G = (V, E)$ with node set $V$, consisting of $N = |V|$ nodes, is associated with a fiber in a bundle at which a force is applied. 
A functional node $i$ breaks or fails if it cannot withstand the force exposed to. 
Its binary state $s_i \in \{0,1\}$ indicates, whether it is functional, $s_i = 0$, or failed, $s_i = 1$. 
The force that is applied to a node is represented by a load $\lambda_i \in \mathbb{R}$ that $i$ carries.
The amount that $i$ cannot withstand anymore is given by its threshold $\theta_i \in \mathbb{R}$.
Usually, $\theta_i$ stays constant, while the load $\lambda_i(t)$ can change in the course of a cascade that involves in discrete time steps $0, \ldots, T$. 
Whenever, a node's load exceeds its threshold, $\lambda_i(t) \geq \theta_i$, $i$ fails.
So, we have  $s_i(t) = H\left(\lambda_i(t) - \theta_i\right)$, where $H(\cdot)$ denotes the Heaviside function. 
The failure of a node impacts its network neighbors that are defined by the link set $E$ of the network.
For instance, in a fiber bundle, the neighboring fibers have to take over the load that was carried by their failed neighbors.  
So, when $i$ fails at time $t$ and it is connected with the still functional node $j$ by a link $(i,j) \in E$, $i$ distributes a load $l_{ij}(t)$ to $j$.
The internal node dynamics are also visualized in Figure \ref{fig:illustrationFailure}. 
With the notation that $l_{ij}(t_{+}) = l_{ij}(t)$ for all later times $t_{+} > t$, we can express the load of a node $j$ as
\begin{align*}
\lambda_j(t+1) =  \sum^N_{i=1} l_{ij}(t)  s_i(t)  + \lambda_j(0),
\end{align*}
where $l_{ij}(t) = 0$ if there is no link $(i,j) \in E$. 
Thus, the failure of a node can cause subsequent failures and trigger a cascade, which ends when no thresholds are exceeded anymore.
In each time step, more than one node can fail, yet, at least one node needs to fail to keep the cascade ongoing. 
So, the dynamics end after maximally $T = N$ steps.

For simplicity, we focus on deterministic dynamics where nodes can fail only once without the option to recover. 
However, since our approach accurately describes the average cascade dynamics in time, recoveries and other form of interventions or changes in time can easily be integrated. 
Similarly, they can be extended by randomizations of failure dynamics. 

Our main cascade size measure on the macro level is the fraction of failed nodes
\begin{align*}
 \rho_N(t) = \frac{1}{N} \sum^N_{i=1} s_i(t).
\end{align*}
When we omit the time dependence, we usually refer to the final cascade size $\rho_N(T)$. 

%
%

\subsubsection{Fiber bundle model}
The main example that we study is a simplified form of fiber bundle model that has been introduced by \cite{moreno2002instability} and further studied by \cite{Lorenz2009} and \cite{Tessone2012} on regular networks.
A node $i$ that fails at time $t$ shares its full load $\lambda_i(t)$ equally among its still functional network neighbors. 
Consequently, we have
\begin{align}
 l_{ij}(t) = A_{ij}(1-s_j(t)) \frac{\lambda_i(t)}{k_i - n_i(t)},
\end{align}
where the adjacency matrix $A$ has elements $A_{ij} = 1$ if $(i,j) \in E$ and $A_{ij} = 0$ otherwise. $k_i = \sum_j A_{ij}$ denotes the degree of node $i$ and $n_i(t) = \sum_j A_{ij} s_{j}$ the number of its failed network neighbors at time $t$. 

This cascade process is Markovian given the full knowledge of all the nodes' loads $\lambda_i(t)$ and thresholds $\theta_i$ at a time $t$. 
The previous history is irrelevant for the determination of $\rho(t+1)$. 
However, the distributed loads $l_{ij}(t)$ are time dependent and summarize this way on the history of the cascade evolution.
It is sensitive to the order of node failures. 
This makes the evolution of the cascade size $\rho_N(t)$ so difficult to solve exactly. 
The key solution is to keep track of the heterogeneous distribution of $l_{ij}(t)$ \emph{before or at time $t$}. 

\subsubsection{Constant load models}
In contrast to fiber bundle models, the load $l_{ij}(t) = w_{ij}$ distributed from a failed to a functional node is constant over time and known apriori.
This makes them insensitive to the precise order of failures and simplifies the local tree approximation of the corresponding cascade size.
For many models of this type, the latter is known for the final cascade size. 
For instance, the description \citep{Gleeson2007} of Watts model \citep{Watts2002} has been inspired by the heterogeneous mean field approximation of Ising models \citep{IsingCascade} and extended to configuration model type random graphs with degree-degree correlations \citep{Dodds2009,Payne2009} or multiplex topologies \citep{BurkholzMultiplex,PhysRevE.91.062813}. 
In similar contexts, several epidemic spreading models have been analyzed \citep{RevModPhys.87.925}.
We show how our framework can still simplify the analytic description of the cascade dynamics in the general case of weighted networks \citep{Burkholz2015} with extension to degree-degree correlation random graphs.

\subsubsection{General dynamics}
Our approach applies to more general cases than threshold models, where the failure of a node is determined by a fixed threshold. 
We can even relax the assumption that a node's state $s_i$ is binary.
We can treat cases where $s_j(t) \in \mathbb{R}$ follows discrete dynamics of the form 
\begin{align}
s_j(t+1) = g\left(\sum^N_{i=1} A_{ij} f(s_i(t), s_j(t), k_i, k_j, \theta_i, \theta_j, w_{ij}, \rho_N(t))\right).
\end{align}
We only require that $g$ and $f$ are measurable functions, where $f$ replaces the failure condition and can depend on the states of the failing as well as load receiving nodes, their degrees, other node attributes $\theta_i, \theta_j$ or link attributes $w_{ij}$ that are allowed to be random (but independent of each other), and even the cascade size $\rho_N(t)$ of the previous time step.
%
%

The arguments that we develop in the following section apply to this case. 
To simplify the derivation however, we focus on the introduced cascade models.
Sec.~\ref{sec:LTA} explains the concept in detail.
A summary of the approach for the specific models is then provided in Sec.~\ref{sec:CL} and Sec.~\ref{sec:LRD}. 

\section{Local tree approximation}\label{sec:LTA}
The cascade processes that we have introduced evolve on a finite and fixed network $G$. 
Yet, we assume that such a network is drawn at random from an ensemble of networks which exhibit a given degree distribution $p(k)$ and degree-degree correlations $p(k,d)$. 
This random graph ensemble is known as extension of the configuration model \citep{Molloy1995,Newman.Strogatz.ea2001Randomgraphswith}, where a neighbor of a node with given degree $k$ has degree $d$ with probability $p(k,d)$ \citep{PhysRevLett.89.208701,PhysRevE.67.026126}.
The original uncorrelated configuration model is a special case for the choice $p(k,d) = p(k)k/z$, where $z$ denotes the average degree $z = \sum_k p(k)k$ in the network.
As this ensemble is maximally random (i.e. it maximizes the entropy of distributions over networks) given the constraints $p(k)$ and $p(k,d)$, it is often assumed as null model for observed phenomena. 
It allows to test the average influence of the degree distribution $p(k)$ and assortativity defined by $p(k,d)$ on processes running on top of such networks.
In the thermodynamic limit of infinite network size $N\rightarrow \infty$, the clustering coefficient vanishes and the network topology becomes \emph{locally tree-like}, if the second moment of the degree distribution $p(k)$ is finite.
This property is essential to calculate the cascade size as average over the random graphs, as the failures of neighbors can be treated as independent.
This independence is sometimes also called \emph{heterogeneous mean field assumption}.
This does not mean that nodes fail independently of each other.
It just acknowledges that in one network realization a node in the network can be connected to an already failed node, while in another it might be connected to a functional one.
The probability of being connected to one neighbor is independent from the probability to be connected to another neighbor.
%

In this configuration model extension, we have constructed undirected networks.
Load can be distributed in both directions of a link-depending on which node fails earlier. 
However, we can introduce a direction by assigning weights $w_{ij}$ and $w_{ij}$ to a link that define a proportions or amounts of load that are distributed along a specific direction.
Some weights can also be set to $w_{ij} = 0$ to obtain directed networks.
Link weights are also drawn initially independently at random from a distribution $p_{W(k_j,k_j)}(w)$ that can depend on the degrees of both involved nodes as introduced in \citep{Burkholz2015}.

Additionally to the network structure, we assume that the nodes' thresholds are drawn initially independently at random from a law that we allow to depend on the degree $k$ of a node. 
We denote the cumulative distribution function of a node with degree $k$ by $F_{\Theta(k)}(\theta)$.
Accordingly, also the initial load that a node carries can be randomly drawn from a distribution $F_{\Lambda_0(k)}(\lambda_0)$.  

In summary, we study an ensemble of random graphs with fixed degree distribution $p(k)$, degree-degree correlations $p(k,d)$, possibly random link weights following $p_{W(k_j,k_j)}(w)$, and random thresholds following $F_{\Theta(k)}(\theta)$ and sometimes random initial loads.
It is quenched in the sense that all quantities are fixed at the beginning of a cascade and do not evolve over time.
The goal is to calculate the cascade size evolution $\rho(t)$ as average with respect to this ensemble in the thermodynamic limit of infinitely large networks.

%
%
%
\subsection{Average cascade size as node failure probability}
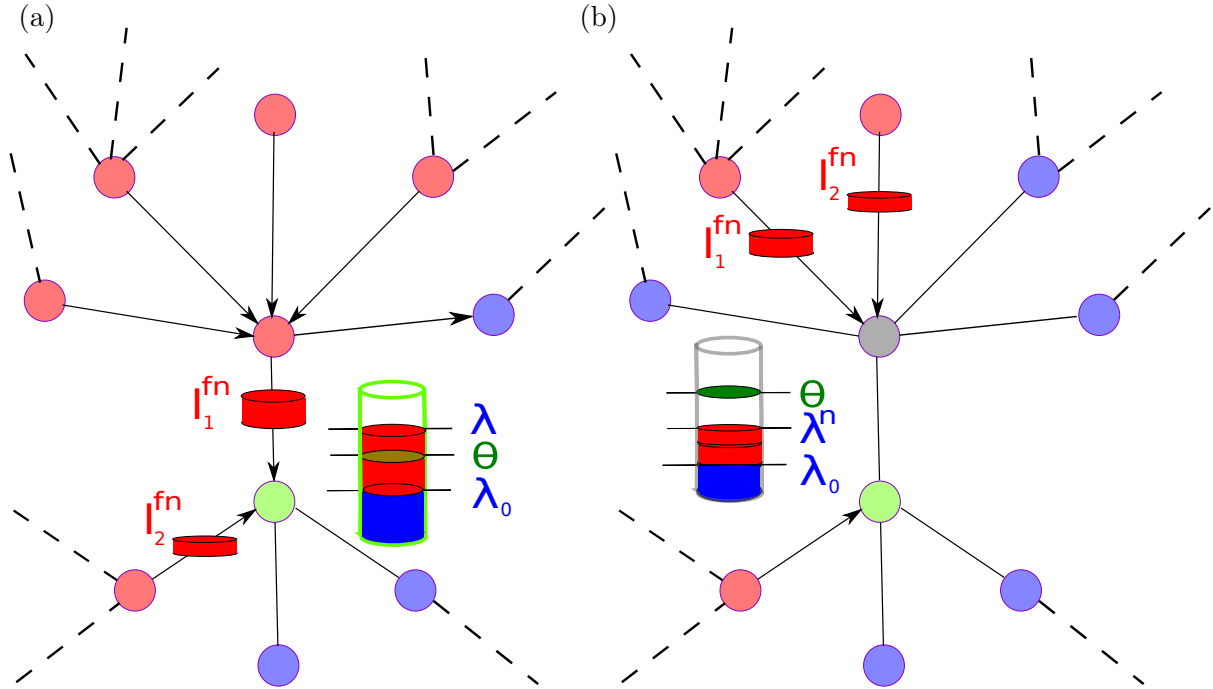
\begin{figure}[t]
 \centering
 \includegraphics[width=0.49\textwidth]{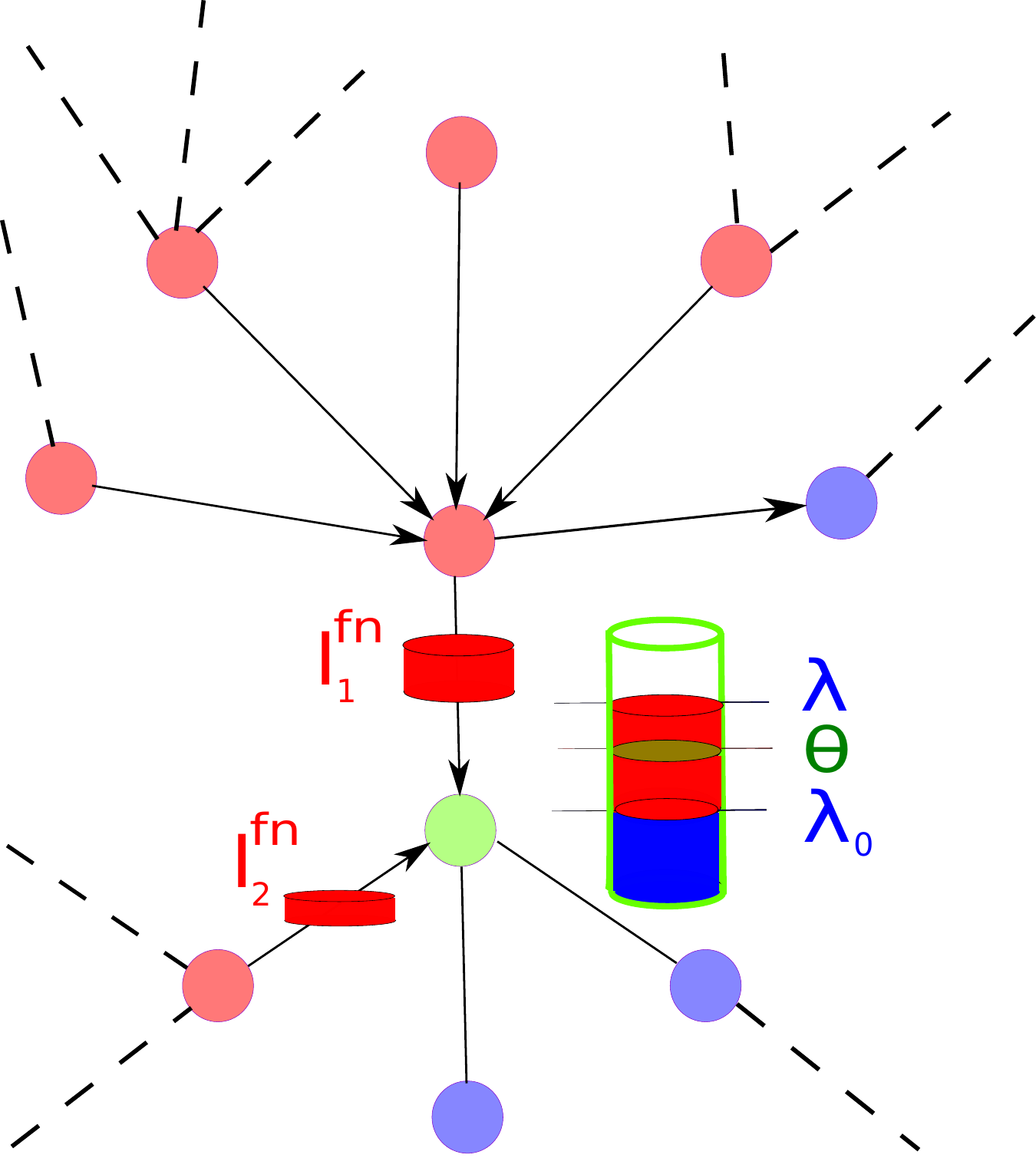}
  \includegraphics[width=0.49\textwidth]{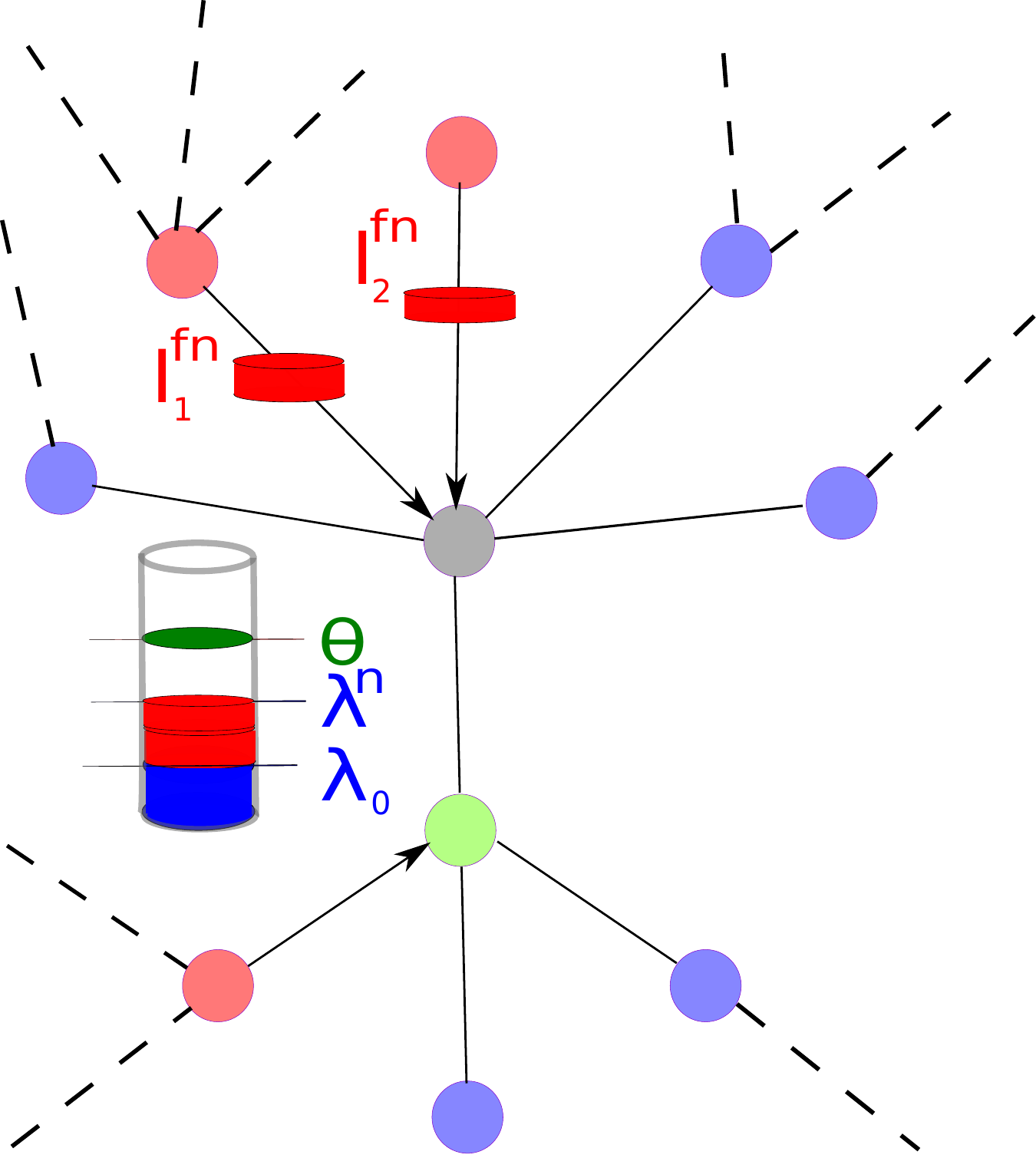}
    \begin{picture}(0,0)
    \put(-450,250){(a)}  
    \put(-240,250){(b)} 
   \end{picture}
 \caption{Illustration of a locally tree-like network structure. Dashed lines lead to other network nodes, which are allowed to be connected. (a) A focal node that is picked uniformly at random from all nodes in the network is colored in green. Realizations of its load $\Lambda = \lambda$ and threshold $\Theta \theta$ are depicted in the green glass. Two failed neighbors have distributed the load $l^{fn}_1$ and $l^{fn}_2$ to the focal node. (b) A focal neighbor is picked uniformly at random from the neighbors of the focal node and is colored in gray. Correspondingly to (a), realizations of the load $\Lambda^n$ and threshold $\Theta^n$ are shown before the failure of the neighbor.}
\label{fig:LocalTree}
\end{figure}

The fraction of failed nodes in an infinitely large network coincides with the probability that a node that we pick uniformly at random from all nodes in the network is failed.
We call such a randomly picked node a focal node, which is visualized in green in Figure~\ref{fig:LocalTree}~(a).
After taking the limit ${N\rightarrow \infty}$, the average cascade size can be written as
\begin{align}
 \rho(t) = \lim_{N \rightarrow \infty} \mathbb{E}\left(\frac{1}{N} \sum^N_{i=1} s^{(N)}_i(t)\right)
 =: \mathbb{P}\left( S(t) = 1\right) = \mathbb{P}\left( \Theta \leq \Lambda(t)\right),
\end{align}
where $S(t)$ denotes the state of a focal node. 
Its failure probability corresponds to the probability that its threshold $\Theta$ exceeds its load $\Lambda(t)$. 
Since this variables are random, we denote them by capital letters, while their realization is written in lowercase.

When the thresholds $\Theta$ and load $\Lambda(t)$ that nodes carry are quite heterogeneous, i.e. they are broadly distributed, mean field approximations that only consider their average cannot lead to accurate results.
Here, we present an iterative procedure to calculate their exact distributions.

To acknowledge that nodes with different degrees can have different failure probabilities, we apply the law of total probability \citep{ProbTheo} and receive
\begin{align}\label{eq:failureCond}
 \rho(t) = \sum^{c}_{k=1} p(k) \mathbb{P}\left( S(t) = 1 \bigm| K = k \right) = \sum^{c}_{k=1} p(k) \mathbb{P}\left( \Theta(k) \leq \Lambda(k,t) \bigm| K = k \right),
\end{align}
where $\mathbb{P}\left( S(t) = 1 \bigm| K = k \right)$ denotes the conditional failure probability of a node given that its degree is $K=k$.
We write $\Lambda(k,t)$ and $\Theta(k)$ to indicate that the distributions of $\Lambda(t)$ and $\Theta$ can depend on the degree $k$ of a node.
The distribution of $\Theta(k)$ is defined by the initial input $F_{\Theta(k)}$.
So, the goal remains to calculate the distribution of the load $\Lambda(k,t)$ that a node with degree $k$ carries at time $t$.

The key in our derivation is the insight that $\Lambda(k,t)$ can be decomposed into a sum of independent random variables
\begin{align}\label{eq:lambda}
 \Lambda(k,t) = \Lambda(k,0) + \sum^k_{j=1} L^{n}_j(k,t-1) = \Lambda(k,0) + \sum^{N_f(k,t-1)}_{j=1} L^{fn}_j(k,t-1),
\end{align}
where $L^{n}_j(k,t-1)$ corresponds to the load that a neighbor has distributed to the focal node \emph{before or at time $t-1$}. 
In case that a neighbor is not failed, we simply have $L^{n}_j(k,t-1) = 0$.
$L^{n}_j(k,t-1)$ are independent because of the locally tree-like network structure and they are equally according to $p_{L^{n}(k,t-1)}(l)$. 
So, their sum is distributed as the $k$-manifold convolution $p^{*k}_{L^{n}(k,t-1)}(l)$, which can be numerically efficiently computed by a Fast Fourier Transformation \citep{Ruckdeschel2010,Frigo2005}.
Thus, we have reduced the problem to finding the distribution of $L^{n}_j(k,t-1)$.

\begin{figure}[t]
 \centering
    \includegraphics[width=0.49\textwidth]{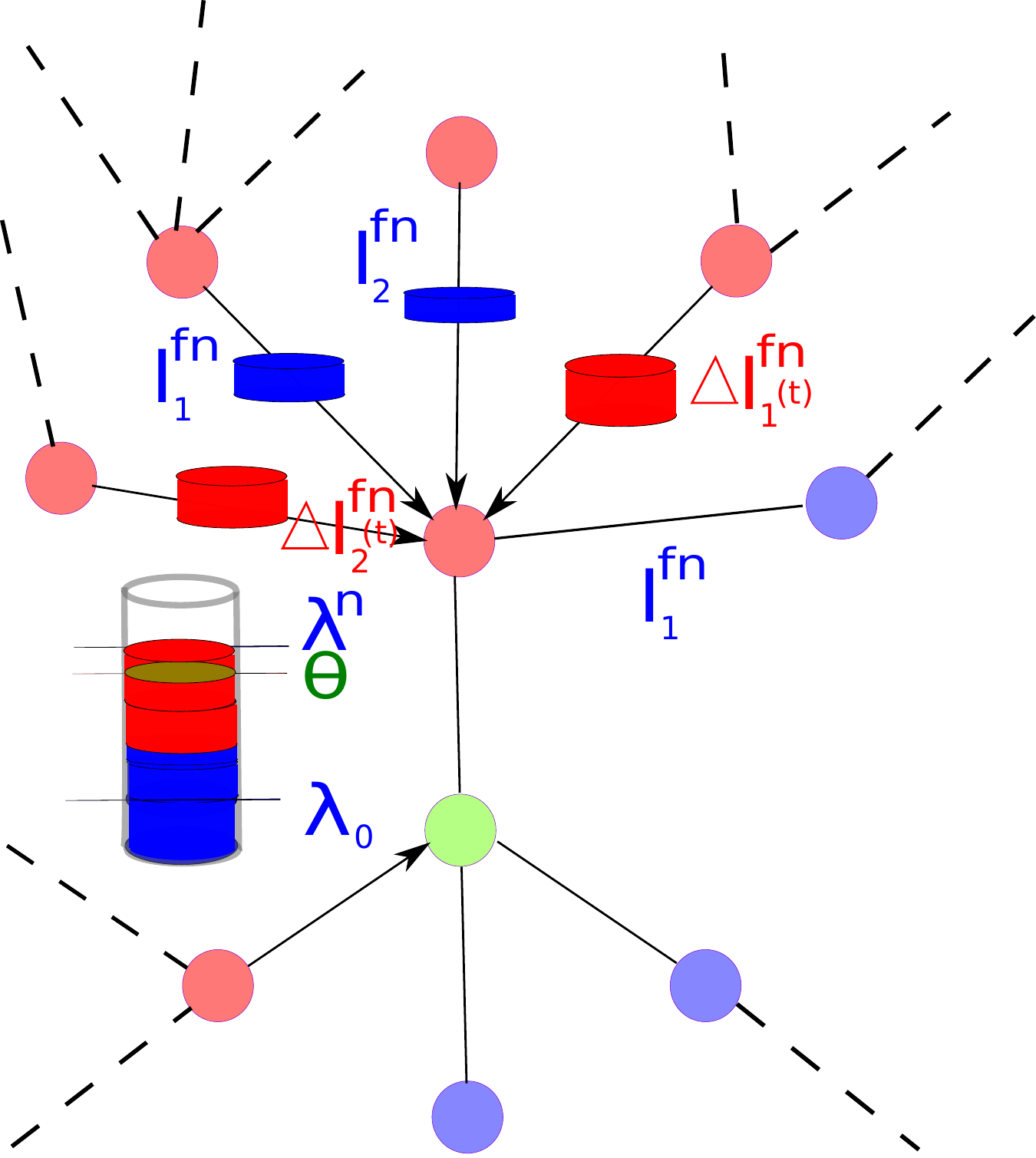}
     \includegraphics[width=0.49\textwidth]{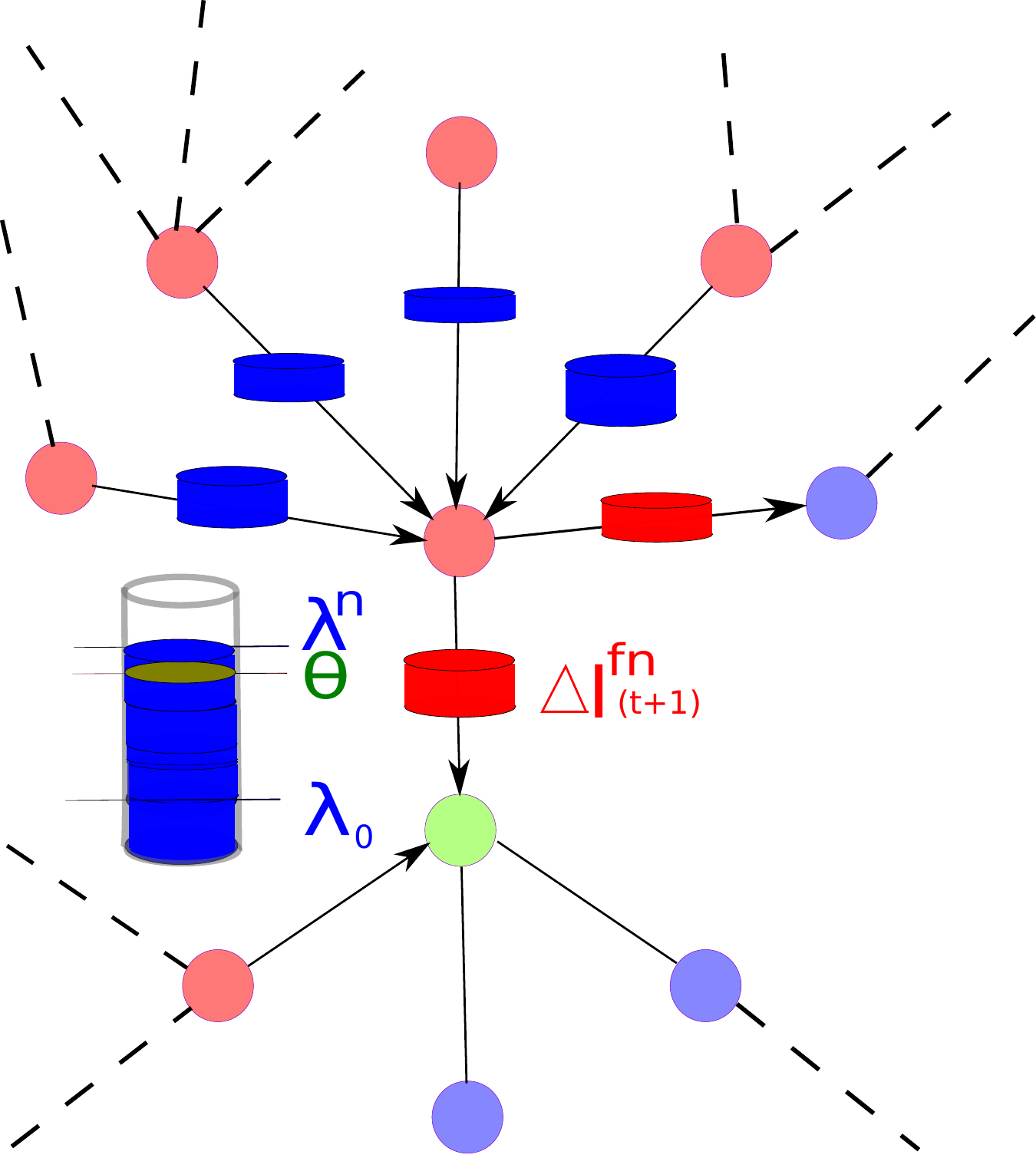}
    \begin{picture}(0,0)
    \put(-450,250){(a)}  
    \put(-240,250){(b)} 
   \end{picture}
 \caption{Illustration of the focal neighbor at failure. (a) The blue load comes from neighbors that have failed before or at time $t-1$ and did not cause the failure of the focal neighbor. The additional red load distributed by actually failing neighbors lead to the failure of the focal neighbor. (b) The focal neighbor distributes the load $\Delta l^{fn}(t+1)$ to the focal node in the next time step.} 
\label{fig:TreeNeighbor}
\end{figure}
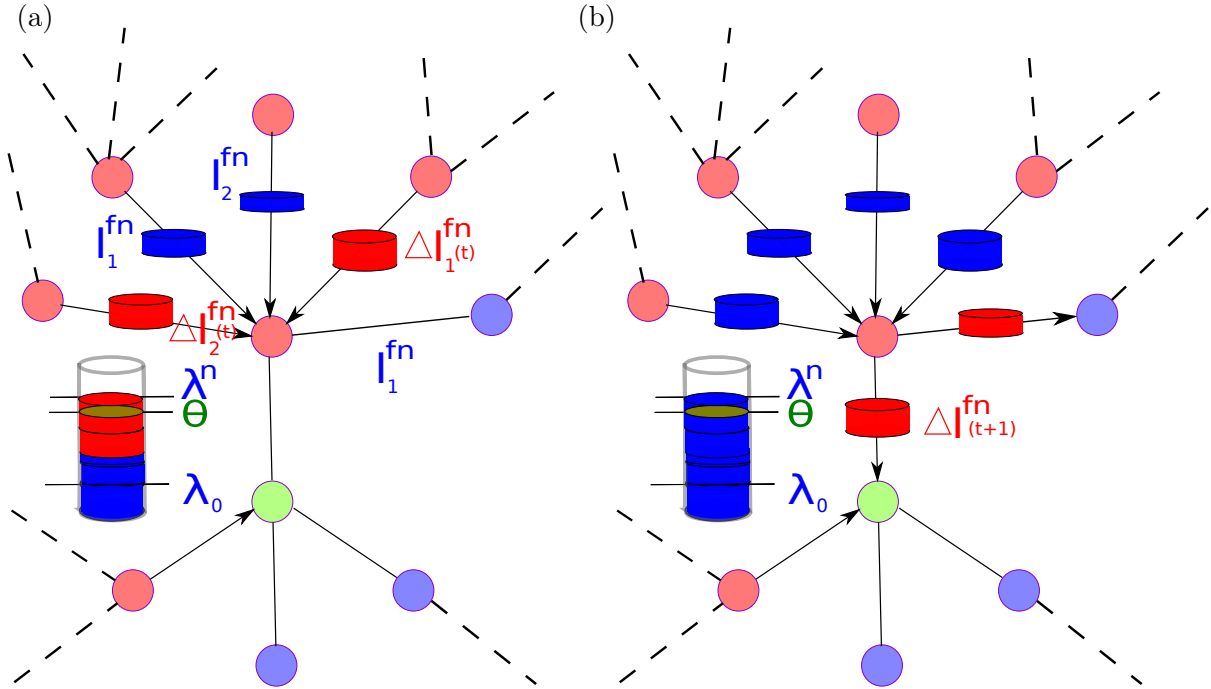
In case of constant load models, this view is very convenient, as the update of the probability distribution of $L^{n}_j(k,t-1)$ is straight forward.

Yet, when we have to keep track of the number $N_f(k,t-1)$ of failed neighbors of a node with degree $k$ (and their time of failure), an employ an alternative view on $\Lambda(k,t-1)$. 
The received load can be expressed as sum over loads $L^{fn}_j(k,t-1)$ distributed by $N_f(k,t-1)$ actually failed neighbors.
There, $L^{fn}_j(k,t-1) = L^{n}_j(k,t-1) ; S^n_j = 1$ is not a normalized random variable.
Instead, $p_{L^{fn}_j(k,t-1)}(l)$ denotes the probability that the neighbor $j$ is failed and distributes the load $l$ to a node with degree $k$. 
It carries the total probability mass
\begin{align}\label{eq:pi}
  \pi(k,t-1) = \mathbb{P}\left(\Theta^n \leq \Lambda^n(t-1)\right) = \int p_{L^{fn}_j(k,t-1)}(l) \; dl. 
\end{align}
that corresponds to the probability that a neighbor of a node with degree $k$ fails before or at time $t$, i.e. the probability that the neighbor's load $\Lambda^n(t-1)$ exceeds its threshold $\Theta^n$.
is independent of the number of neighbors $N_f(k,t-1)$ that have failed before or at time $t-1$. 
We only have to respect that the focal neighbor's degree $D$ is distributed by $p_D(d) = p(k,d)$ instead of $p(k)$.
Furthermore, for the evaluation of the failure condition of a focal node in Eq.~\ref{eq:failureCond}, we are only interested in the case that the neighbor has failed before the focal node so that it can cause its failure.
Consequently, only the remaining $D-1$ neighbors of the focal neighbor can have led to the failure of the focal neighbor.
So, we have:
\begin{align}
 \pi(k,t-1) = \sum_d p(k,d) \mathbb{P}\left(\Theta(d) \leq \Lambda(d,0) + \sum^{d-1}_{j=1} L^{n}_j(d,t-2)\right).
\end{align}
Clearly, also the number of failed neighbors $N_f(k,t-1)$ depends on $\pi(k,t-1)$.  
Because of the locally tree-like network structure, $N_f(k,t-1)$ is binomial distributed $N_f(k,t-1) \sim B\left(k, \pi(k,t-1)\right)$,

In summary, Eq.~\ref{eq:failureCond} can be written as
\begin{align}\label{eq:rhot}
 \rho(t) = \sum_k p(k) \sum^k_{f=0}\binom{k}{f} \left(1-\pi(k,t-1)\right)^{k-f} \int p_{\Lambda_0}*p^{*f}_{L^{fn}(k,t-1)}(l) F_{\Theta(k)}(l) \; dl.
\end{align}
where $p_{\Lambda_0}*p^{*f}_{L^{fn}(k,t-1)}$ is the density of $\Lambda(k,0) + \sum^f_{j=1} L^{fn}_j(k,t-1)$, $F_{\Theta(k)}$ the cumulative distribution function of a node's threshold, and $\pi(k,t-1)$ the failure probability of a neighbor of a node with degree $k$.

\subsection{The load distributed by a failed neighbor}


When the load $L^{fn}_j(k,t-1)$ that a failed neighbor distributed \emph{before or at time $t-1$} changes in time, we have to successively update it by the load $\Delta L^{fn}_j(k,t)$ that a neighbor failing \emph{exactly at time $t$} distributes:
\begin{align}
 p_{L^{fn}(k,t)}(l) = p_{L^{fn}(k,t-1)}(l) + p_{\Delta L^{fn}(k,t)}(l).
\end{align}
For $l > 0$, we also have $p_{L^{n}(k,t)}(l) = p_{L^{n}(k,t-1)}(l) + p_{\Delta L^{fn}(k,t)}(l)$.
Fiber bundle models, for instance, are challenging, because the load that a failing neighbor distributes $\Delta L^{n}_j(k,t)$ at time $t$ depends on the load that this neighbor carries and its number of functional (surviving) neighbors $N_{s}$ at the time of its failure: $\Delta L^{fn}(k,t) = f\left(\Lambda^{n}(t), N_{s}(t-1)\right)$.

To determine $p_{\Delta L^{fn}(k,t+1)}(l)$, we thus have to look at the situation at the time of failure of the focal neighbor, which is illustrated in Figure \ref{fig:TreeNeighbor}.
Let's assume, the focal neighbor has degree $D = d$.
Before its failure at time $t+1$, $N_o$ of its neighbors have already failed before or at time $t-1$ without causing the failure of the focal neighbor, see also Figure~\ref{fig:LocalTree}~(b).
The focal neighbor carries the load
\begin{align}
 \Lambda^{n}(d,t) =\Lambda^{n}(d,0) + \Lambda_{o} = \Lambda^{n}(d,0) + \sum^{N_o}_{j=1} L^{fn}_j(d,t-1),
\end{align}
which fulfills the constraint $\Lambda^{n}(d,t) < \Theta^n(d)$, as shown in Figure~\ref{fig:TreeNeighbor}~(a). 
In the next time step $t$, $N_n$ additional neighbors fail so that the focal neighbor receives the load $\sum^{N_n}_{i=1} \Delta L^{fn}_j(d,t)$.
This additional load causes now the failure of the focal neighbor, as visualized in Figure~\ref{fig:TreeNeighbor}~(a).
Furthermore, $N_a$ of the neighbors might fail at exactly the same time as the focal neighbor so that $N_s + 1 = D-N_o-N_n-N_a$ nodes receive load by the failing focal neighbor. 
The failing focal neighbor distributes then the load:
\begin{align}\label{eq:DeltaL}
 \Delta L^{fn}(k,t+1) = f\left(\Lambda^{n}(D,0) + \sum^{N_o}_{j=1} L^{fn}_j(D,t-1) + \sum^{N_n}_{i=1} \Delta L^{fn}_i(D,t), N_s \right)
\end{align}
to the focal node, where the variables respect the constraint $\Lambda^{n}(D,t) < \Theta^n(D) \leq \Lambda^{n}(D,t+1)$.
The variables 
\begin{align}
 (N_o,N_n,N_a,N_s) \sim M\left(d-1,\pi(d,t-1),\pi(d,t)-\pi(d,t-1),\pi(d,t+1)-\pi(d,t),1-\pi(d,t+1)\right)
\end{align}
follow a multinomial distribution for $D = d$.
This information determines the iterative update of the cascade size $\rho(t)$ according to Eq.~(\ref{eq:rhot}).
We summarize and specify the approach for two examples, i.e. the introduced fiber bundle and constant load models.

\subsection{Constant load models on weighted network ensembles}\label{sec:CL}
We start with the simpler case of Constant load models, where load that a neighbor with degree given degree $d$ distributes to a node with degree $k$ does not depend on the cascade history. 
Instead it is defined apriori by link weights following the probability distribution $p_{W(d,k)}(w)$.
The following local tree approximation is based on Eq.~(\ref{eq:lambda}) and extends the approach by \cite{Burkholz2015} to capture degree-degree correlations.
The treated special case of uncorrelated networks is simplified algorithmically here as well.

Initially, all nodes (and neighbors) with thresholds $\Theta \leq 0$ fail with probability
\begin{align}
 \rho(0) = \sum_k p(k) F_{\Theta(k)} (0), \ \ \ 
 \pi(k,0) = \sum_d p(k,d) F_{\Theta(d)} (0)
\end{align}
and distribute the load
\begin{align}
\begin{split}
   p_{L^{n}(k,t)}(0) = & \left( 1-\pi(k,0)\right) +
 \sum_d p(k,d) F_{\Theta(d)} (0) p_{W(d,k)}(0),
\\
p_{L^{n}(k,t)}(l) = & \sum_d p(k,d) F_{\Theta(d)} (0) p_{W(d,k)}(l) \text{ for } l \neq 0.\\
\end{split}
\end{align}
Then, each further time step $t+1$ follows from $t$ by the update of the load $L^n(k,t)$ that is distributed to a node with degree $k$ by a neighbor.
The full dynamics are captured by
\begin{align}\label{eq:LnCL}
\begin{split}
\mathbb{P}\left(S^n(t) = 1 \big| D = d\right) = & \mathbb{P}\left(\Theta(d) \leq \sum^{d-1}_{j=1} L^n_j(d,t-1)\right) = \int p^{*(d-1)}_{L^n(d,t-1)}(l)  F_{\Theta(d)}(l)\; dl,\\
\pi(k,t) = & \sum_d p(k,d) \mathbb{P}\left(S^n = 1 \big| D = d\right),\\
 p_{L^{n}(k,t)}(0) = & \left( 1-\pi(k,t)\right)  
 + 
 \sum_d p(k,d) \mathbb{P}\left(S^n(t) = 1 \big| D = d\right) p_{W(d,k)}(0),
\\
p_{L^{n}(k,t)}(l) = & \sum_d p(k,d) \mathbb{P}\left(S^n(t) = 1 \big| D = d\right) p_{W(d,k)}(l) \text{ for } l \neq 0,\\
\rho(t+1) = & \sum_k p(k) \int p^{*k}_{L^n(k,t)}(l)  F_{\Theta(k)}(l)\; dl.  
\end{split}
\end{align}
To simplify the notation, we compute explicitly the failure probability $\mathbb{P}\left(S^n(t) = 1 \big| D = d\right)$ of a focal neighbor with degree $d$ and the failure probability $\pi(k,t)$ of a focal neighbor that is connected to a node with degree $k$. 
Then, a neighbor does not distribute any load ($L^{n}(k,t) = 0$), when it is either not failed (with probability $1-\pi(k,t)$) or is failed but the link weight is zero ($W(d,k) = 0$).
It has degree $d$ with probability $d$ and is additionally failed with probability $\mathbb{P}\left(S^n(t) = 1 \big| D = d\right)$.
In this case, it distributes the load $l$ with probability $p_{W(d,k)}(l)$.
This fully determines the load distribution, which enables the calculation of the cascade size according to Eq.~(\ref{eq:failureCond}) and Eq.~(\ref{eq:lambda}).

\subsection{Fiber bundle model}\label{sec:LRD}
The fiber bundle model requires the knowledge of the load $\Lambda^n(k,t)$ that a focal neighbor of a node with degree $k$ carries, as this is distributed among its $D-N_o-N_n-N_a$ functional network neighbors at the time of load distribution. 
Precisely, Eq.~(\ref{eq:DeltaL}) specializes to
\begin{align}
 \Delta L^{fn}(k,t+1) = \frac{\Lambda^{n}(D,0) + \sum^{N_o}_{j=1} L^{fn}_j(D,t-1) + \sum^{N_n}_{i=1} \Delta L^{fn}_i(D,t)}{D-N_o(k,t-1)-N_n(k,t)-N_a(k,t+1)}
\end{align}
with $\Lambda^n(k,t) < \Theta^n(D) \leq \Lambda^n(k,t+1)$.

Let's assume that all nodes receive the same deterministic initial load $\lambda_0$. 
We summarize the iterative calculation of the cascade dynamics for the fiber bundle model for infinitely large degree-degree correlated configuration model type random graphs.

Initially, we start from the probability distribution of the load that is distributed by initially failing neighbors:
\begin{align}
\begin{split}
 \pi(k,0) & = \sum_d p(k,d) F_{\Theta(d)}(\lambda_0), \ \ \ \ p_{\Delta L^{fn}(k,0)}(0)   = 1 - \pi(k,0),\\
p_{\Delta L^{fn}(k,0)} \left(\frac{\lambda_0}{n_s+1} \right) & = \sum^c_{d= n_s + 1} p(k,d) \binom{d-1}{n_s} (\pi(k,0))^{d-1-n_s} \left(1-\pi(k,0)\right)^{n_s} F_{\Theta(k)}(\lambda_0)
\end{split}
\end{align}
 for $n_s = 0, \cdots, c-1$, where $c$ denotes the maximal degree in the network or, if this does not exist, $c = \infty$.  
Just formally, we define the probability distribution of the load that is distributed by nodes before time $t=0$ as zero: $\mathbb{P}\left(L^{n}(k,-1)= 0 \right) =  1$ and $p_{L^{fn}(k,-1)}(l) =  0$.
On this basis, we can construct all following time steps iteratively.

We first determine the failure probability of a neighbor at time $t+1$ as:
\begin{align*}
\pi(k,t+1) = \sum_d p(k,d) \int p^{*(d-1)}_{L^n(d,t)}(l)  F_{\Theta(d)}(l)\; dl.
\end{align*}
and compute next the distribution of load by a neighbor exactly at time $t+1$.
In the following $n_o$ denotes the number of previously failed neighbors (before or at time $t-1$), $n_n$ the number of neighbors that have failed at time $t$, $n_a$ the number of neighbors that fail at time $t+1$ like the focal neighbor and $n_s$ the remaining functional neighbors of the $d-1$ possibly failing neighbors.
We denote with $I_{d-1} = \left\{ \underline{n} = (n_o,n_n,n_a,n_s) \in \{0,\ldots,d-1\}^4 \big| n_o + n_n + n_a + n_s = d-1\right\}$ the index set of these values so that their sum equals $d-1$.
Recall that $F_{\Theta(d)}$ is the threshold distribution of a focal node or neighbor with degree $d$.
We obtain:
\begin{align*}
  p_{\Delta L^{fn}(k,t+1)}( l)  = & \sum_{\underline{n} \in I_{d-1}}\frac{(d-1)!}{n_o! n_n! n_a! n_s!} \left(1-\pi(t+1)\right)^{n_s}\left(\pi(t+1)-\pi(t)\right)^{n_a}  \int^{\infty}_{x=0}p^{*n_n}_{\Delta L^{fn}(d,t)}[x] \\ 
 & \times  p^{*n_o}_{L^{fn}(d,t-1)}[l(n_s + 1) - \lambda_0 - x] \left(F_{\Theta(d)}\left(l (n_s + 1)\right) - F_{\Theta(d)}\left(l (n_s + 1) - x\right) \right) \; dx
\end{align*}
for $l>0$. 
In total, a failing neighbor carries here the load $l(n_s + 1) = \lambda_0 + z + x$ at failure, where it received $z = l(n_s + 1)-x$ because of neighboring failures before or at time $t-1$.
This has not been enough to exceed the threshold $\theta \in {]l(n_s + 1)-x ,l(n_s + 1)]}$ of the focal neighbor. 
Yet, the additional load $x$ received at time $t$ causes the failure of the focal neighbor, which defines the upper bound of possible threshold values of the neighbor.

With this information, we can update the density of the load $L^{n}(k,t)$ that is distributed by a neighbor before or at time $t$ to $L^{n}(k,t+1)$ by
\begin{align*}
 p_{L^{n}(k,t+1)}(l) =  p_{L^{n}(k,t)}(l) + p_{\Delta L^{fn}(k,t+1)}(l) \ \ \ \ \ \text{for all } l \in \mathbb{R}\backslash\{0\}
\end{align*}
and $p_{L^{n}(k,t+1)}(0) = 1- \pi(t+1)$. 
This determines the fraction of failed nodes $\rho(t+2)$ already one time step later:
\begin{align}
 \rho(t+2)  = \sum_k p(k) \mathbb{P}\left(\Theta(k) \leq \sum^k_{j=1} L^{n}_j(k,t+1) \right) = \sum_k p(k) \int^{\infty}_{0} F_{\Theta(k)}(x) p^{*k}_{L^{n}(k,t+1)}(x) \; dx.
\end{align}
Despite the fact that the involved load distributions are discrete, we approximate the distributions on an equidistant grid in our numerical calculations. 
This allows us to compute the convolutions of load distributions by Fast Fourier Transformation. 
The integrals simplify to actual sums over discrete values of distributed load.

%
\section{Comparison of numerical local tree approximations with simulation results}

%
%

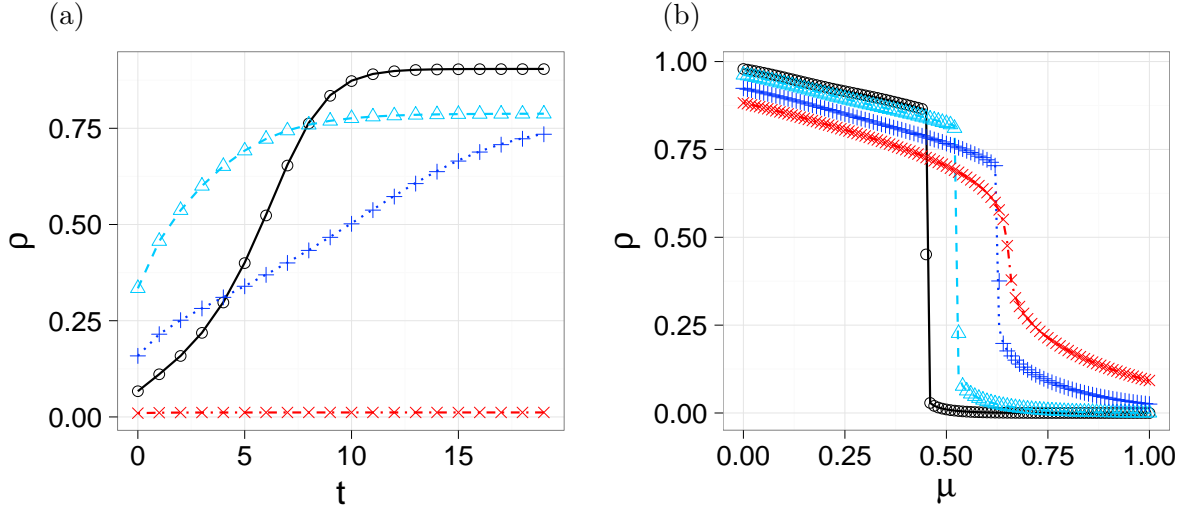
\begin{figure}[t]
 \centering
 \includegraphics[width=0.49\textwidth]{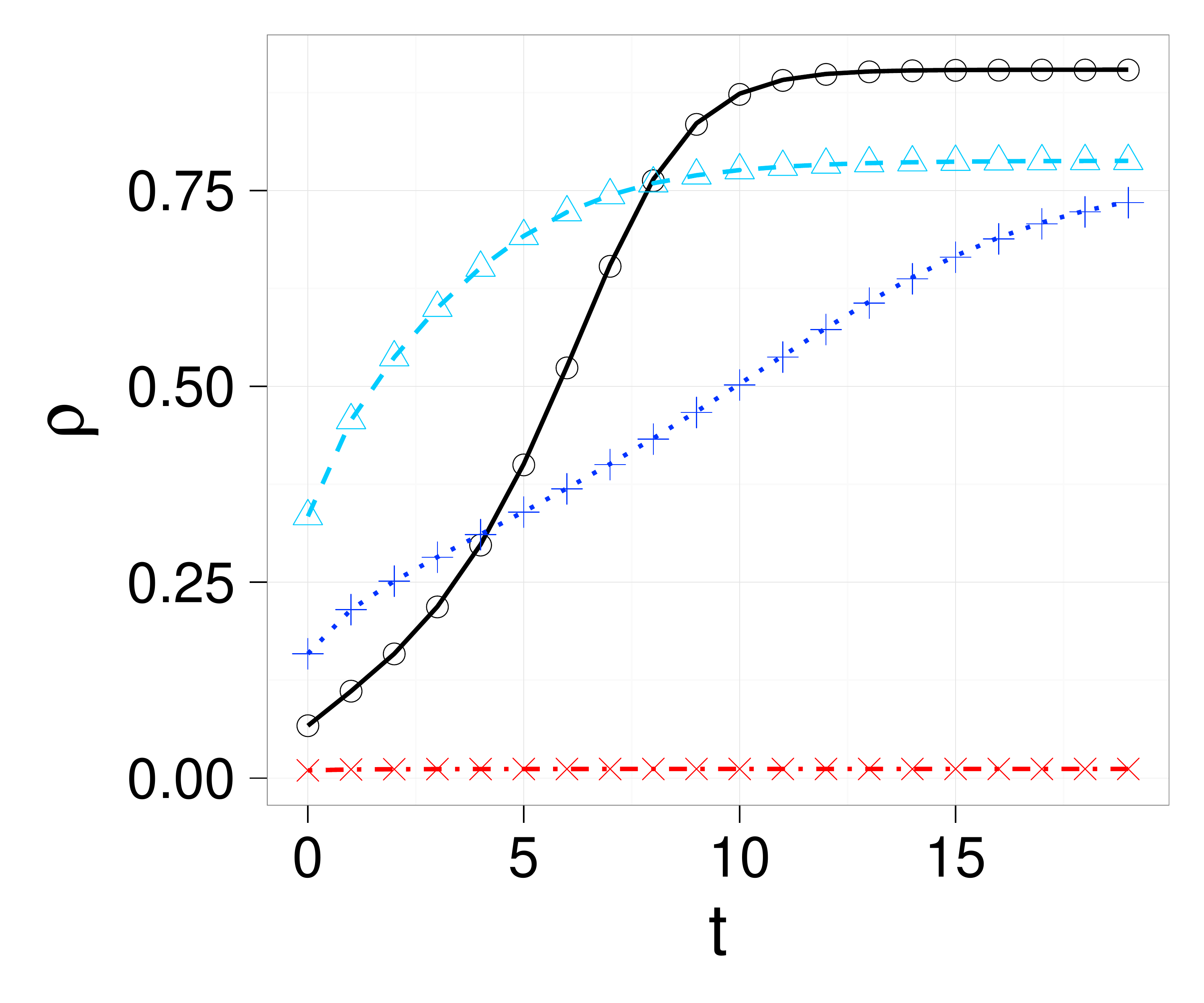}
\includegraphics[width=0.49\textwidth]{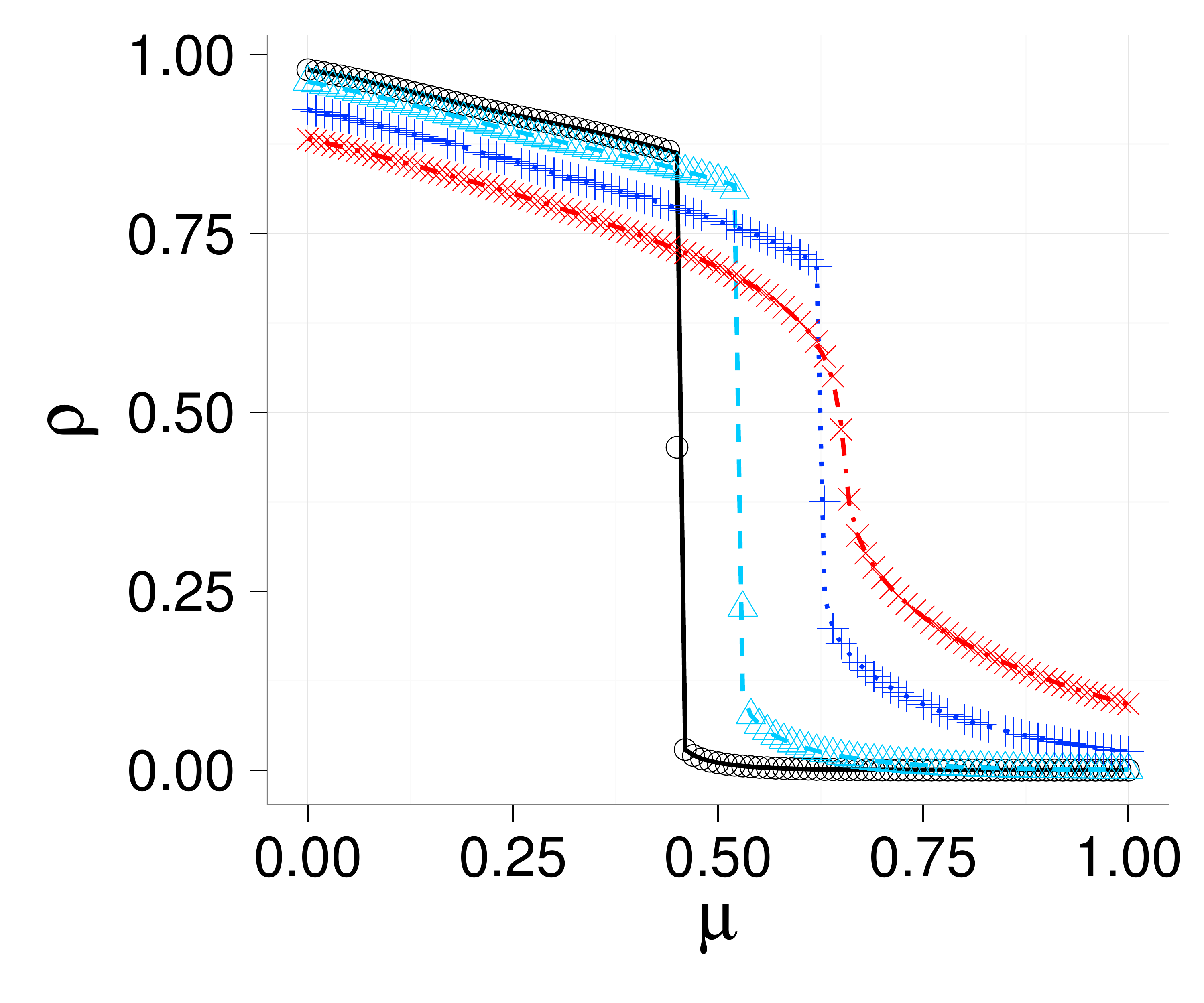}
\begin{picture}(0,0)
    \put(-430,190){(a)}
    \put(-200,190){(b)}
   \end{picture}
 \caption{Comparison of numerical local tree approximations (LTA) and simulations for the fiber bundle model and Poisson random graphs with average degree $z=3$, where lines represent the LTA and symbols in the same color correspond to simulation results. 
 The thresholds $\Theta$ are normally distributed with mean $\lambda_0 + \mu$ and standard deviation $\sigma$ ($\Theta \sim \mathcal{N}(\lambda_0 + \mu, \sigma^2)$). The initial load for all nodes is $\lambda_0 = 0.5$.
 (a) We show the cascade size evolution.
 Black circles belong to $(\mu, \sigma) = (0.3, 0.2)$, dark blue plus signs $+$ to $(\mu, \sigma) = (0.5, 0.5)$. Light blue triangles depict $(\mu, \sigma) = (0.3, 0.7)$, while red and the symbol $x$ belong to $(\mu, \sigma) = (0.7, 0.3)$.
  (b) Final cascade size after $T=300$ fixed point iterations. Black circles belong to $\sigma = 0.2$ and
  light blue triangles to $\sigma = 0.3$.
  Dark blue plus signs $+$ depict $\sigma = 0.5$, while red and the symbol $x$ belong to $\sigma = 0.7$.
  }
\label{fig:LRD_AnaSim}
\end{figure}


As proof of concept, we compare the results of numerical local tree approximations (LTA) with Monte Carlo simulations on uncorrelated Poisson random graphs with degree distribution $p(k) = e^{-z} z^k/k!$ and $p(k,d) = p(d)d/z$.
For uncorrelated networks, the numerical iteration simplifies, as the failure probability of a neighbor $\pi(k,t)$ and the distributed load $\Delta L^{fn}(k,t)$ become independent of the degree $k$ of the load receiving node.
In this set-up, our LTA requires negligible computational resources and converges in the range of a few minutes (usually less than a minute), while simulations often require several hours.

Our simulations calculate the average cascade size over $500$ independent realizations of networks consisting of $N = 10^5$ nodes.
As \citep{Lorenz2009,Tessone2012}, we assume that all nodes are equipped with the same initial load $\lambda_0$ and that the thresholds follow a normal distribution independent of the degree of a node, i.e. $\Theta \sim \mathcal{N}\left(\mu + \lambda_0, \sigma^2\right)$. 

%
%
Two constant load models have been analyzed by \citep{Burkholz2015} in the described set-up. 
As the results of the described LTA coincide with the approach presented there, we focus here on fiber bundle models. 
Fig.~\ref{fig:LRD_AnaSim} shows perfect agreement between our simulations and our LTA. 
The full cascade size evolution is captured as indicated by Fig.~\ref{fig:LRD_AnaSim}(a). 
The convergence speed to the final state and the general shape of the cascade evolution can differ substantially for different threshold parameters. 
Interestingly, most failures happen only after a few time steps. 
Fig.~\ref{fig:LRD_AnaSim}(a) compares the final cascade sizes for several threshold parameters. 
Also the sharp regime shifts are accurately computed by our numerical calculations. 

\begin{figure}[h]
 \centering
 \includegraphics[width=0.32\textwidth]{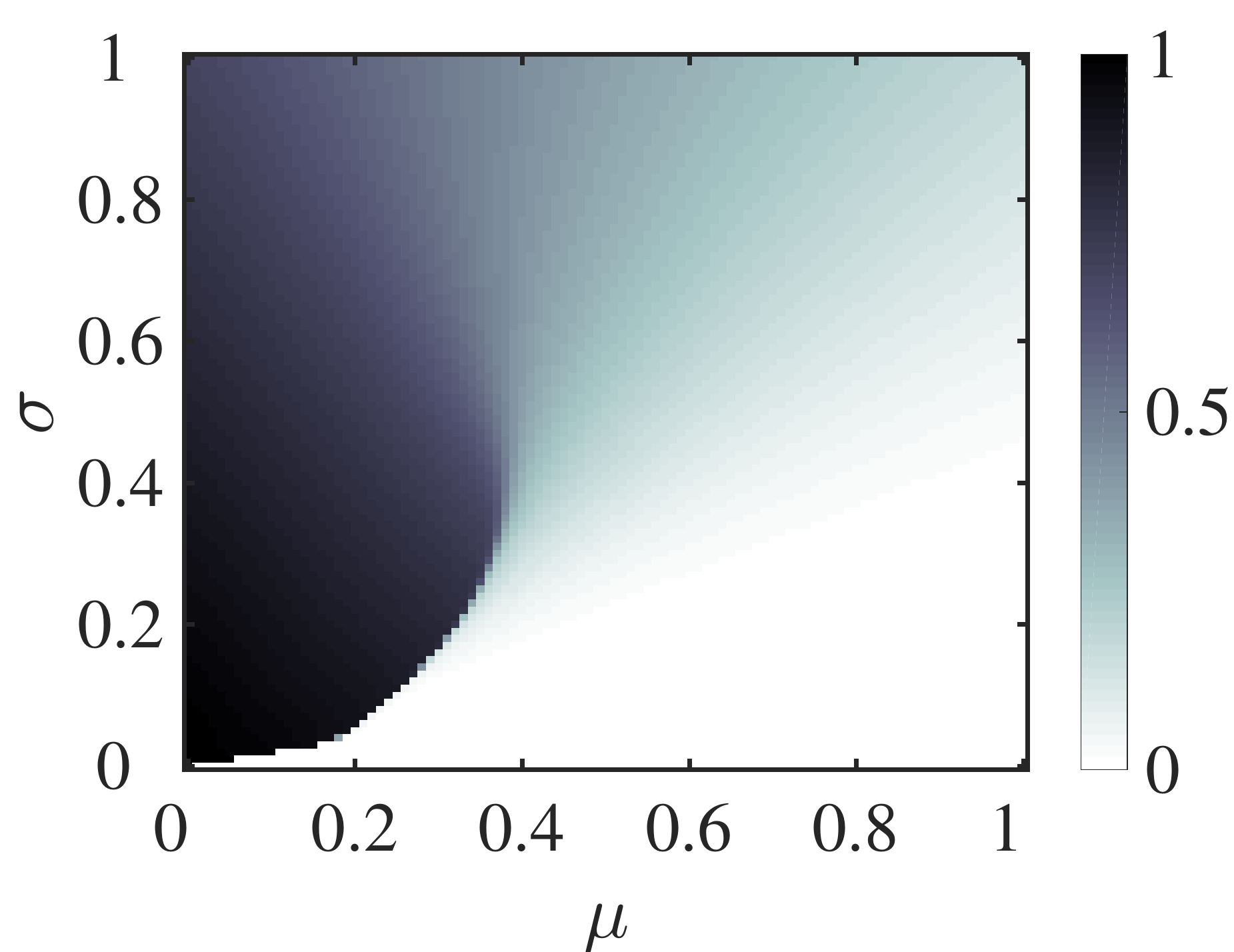}
\includegraphics[width=0.32\textwidth]{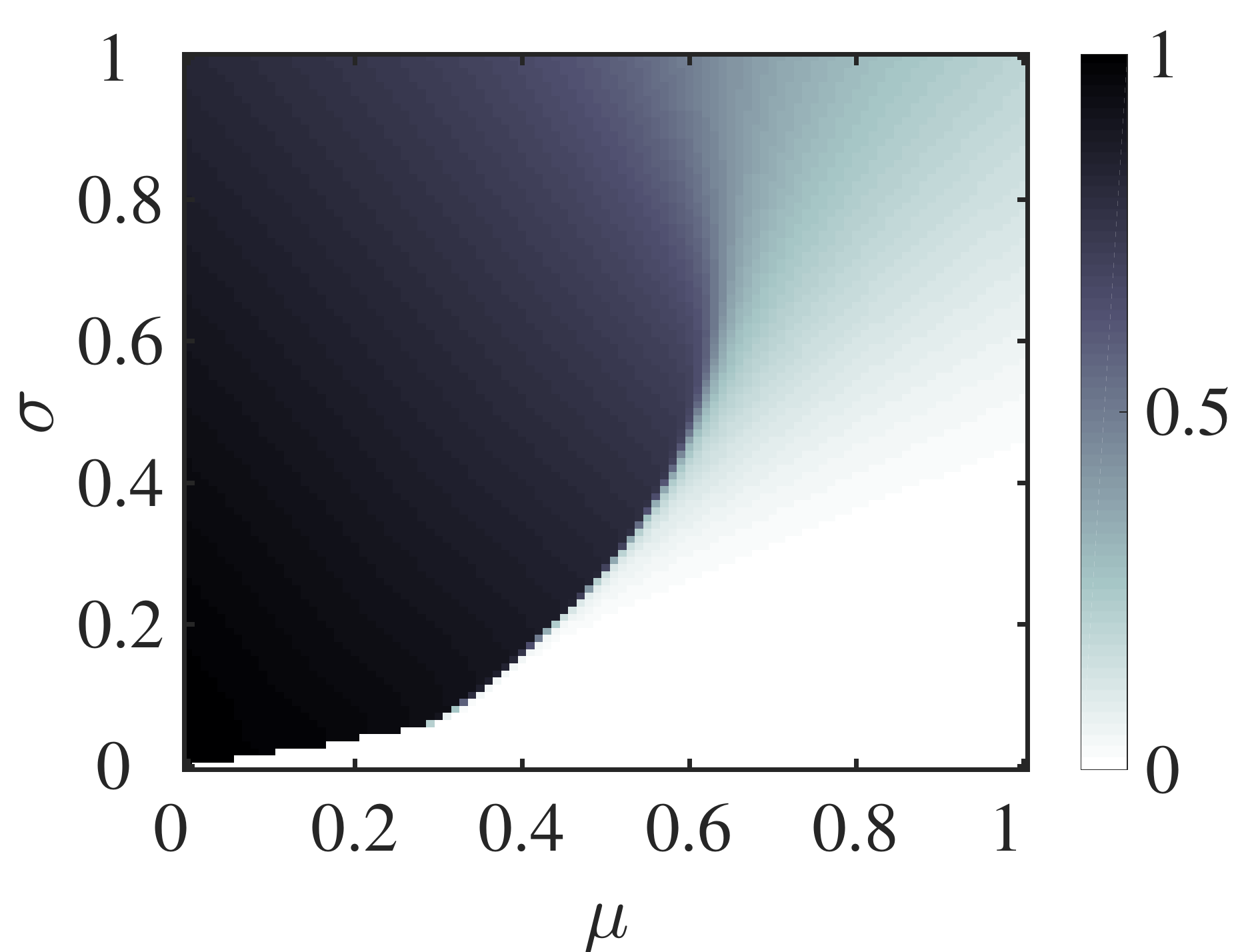}
 \includegraphics[width=0.32\textwidth]{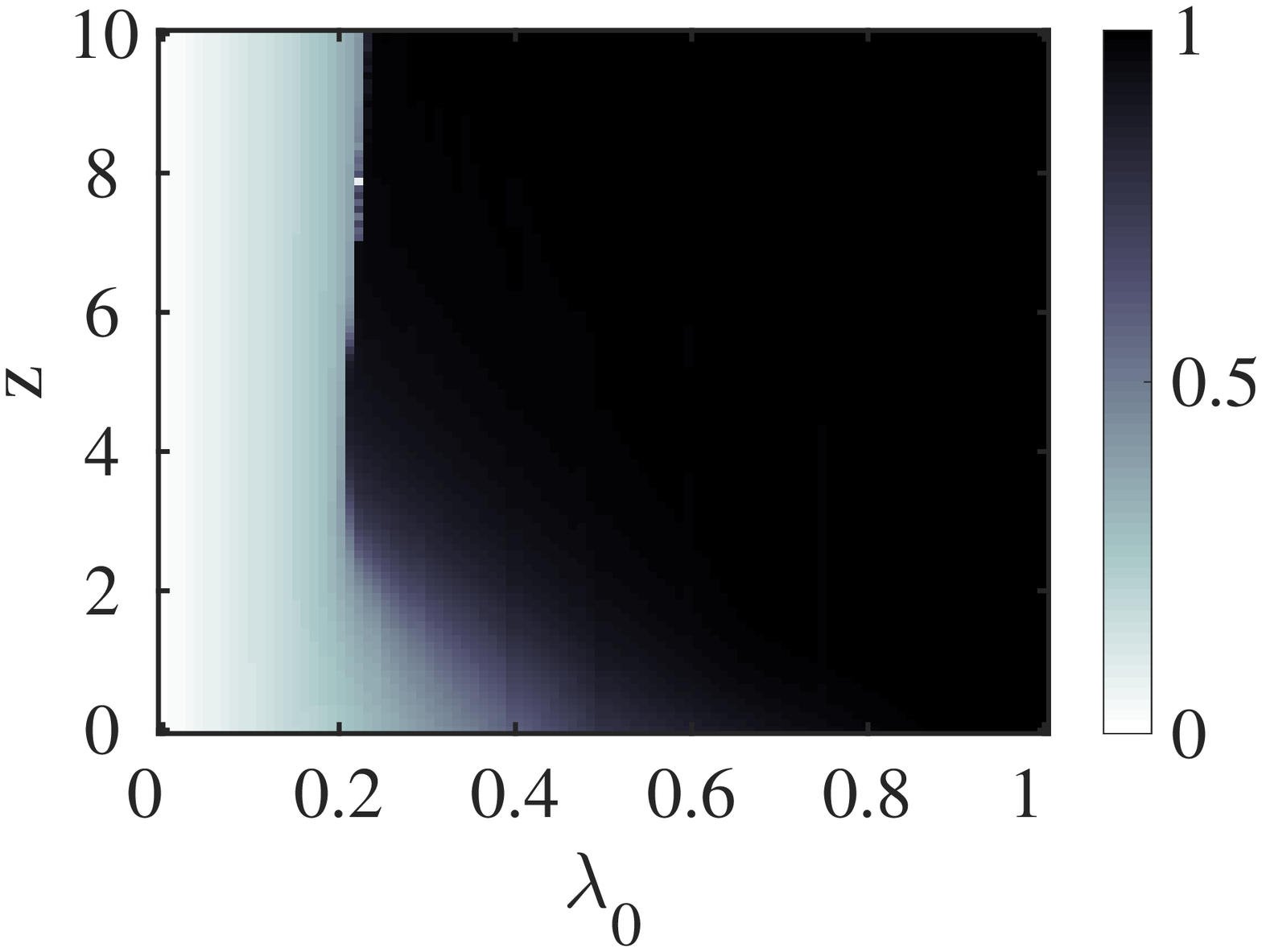}
\begin{picture}(0,0)
    \put(-445,115){(a)}
    \put(-295,115){(b)}
    \put(-145,115){(c)}
   \end{picture}
 \caption{Final fraction of failed nodes $\rho$ for the fiber bundle model after $50$ fixed point iterations for Poisson random graphs (with average degree $z$) obtained by an LTA. $\rho$ is color coded. The darker the color, the bigger is the cascade size. Each node receives the initial load $\lambda_0$. The thresholds $\Theta_i$ are independently distributed.
(a-b) Normally distributed thresholds with mean $\mu + \lambda_0$ and standard deviation $\sigma$ ($\Theta \sim \mathcal{N}(\mu + \lambda_0, \sigma^2)$). 
(a) $\lambda_0 = 0.3$ , (b) $\lambda_0 = 0.5$.
(c) Uniform threshold distribution $\Theta \sim U[0,1]$. 
}
\label{fig:LRD_Ana} 
\end{figure}

For completeness, we provide several phase diagrams for the final cascade size to give an overview of the typical influence of the model parameters. 
Fig.~\ref{fig:LRD_Ana}~(a-b) show normally distributed thresholds for two different values of the initial load $\lambda_0$.
This extra degree of freedom, $\lambda_0$, in the model clearly influences the size of the region of big cascade sizes, but does not lead to a qualitative change of its form. 
As for fully connected networks \citep{Lorenz2009}, we observe a sharp regime shift for decreasing average threshold $\mu + \lambda_0$ and mediocre values of the standard deviation $\sigma$. 
Interestingly, for increasing $\sigma$ the final (average) cascade size declines again for mediocre values of $\mu$.    
We have also tested uniformly distributed thresholds, as they are often considered in fiber bundle models in the literature \citep{RevModPhys.82.499}.
Fig.~\ref{fig:LRD_Ana}~(c) explores the role of the initial load $\lambda_0$ and the average degree $z$ for uniformly distributed thresholds. 
For big enough initial load $\lambda_0$, almost the whole system breaks down, while $\rho$ is negligibly small for for small enough $\lambda_0$.
The average degree $z$ influences the outcome only in a small region of $\lambda_0$. 

In addition to the cascade size evolution, the LTA provides interesting further information about the time of failure of nodes conditional on their degree. 
This allows to analyze, for instance, which nodes are the main spreaders of failures and when they fail in the course of a cascade.
Interventions to reduce or enhance the cascade size can use this information to target specific nodes. 
In fiber bundle models, the goal is usually to prevent further failures.  
This can, for instance, either be achieved by saving nodes whose failure would cause many subsequent failures or to force them to fail early before they can accumulate a high amount of load to spread to their neighbors in case of a later failure.

Both, the failure probability exactly at time $t$ of a node conditional on its degree $k$ and the probability that a node fails at time $t$ and has degree $k$ give insights about possible intervention strategies. 
The conditional failure probability corresponds to the nodes with degree $k$ that fail exactly at time $t$ as fraction over all nodes with degree $k$ in the infinitely large network: 
 \begin{align}
 \begin{split}
  \mathbb{P}\left(S(t) = 1 \big| K = k\right) & - \mathbb{P}\left(S(t-1) = 1 \big| K = k\right) \\
 & =  \mathbb{P}\left(\Theta(k) \leq \sum^{k}_{j=1} L^n_j(d,t-1)\right) -  \mathbb{P}\left(\Theta(k) \leq \sum^{k}_{j=1} L^n_j(d,t-2)\right). 
 \end{split}
  \end{align}
The probability that a node fails at time $t$ and has degree $k$
 \begin{align}
 \begin{split}
  \mathbb{P}\left(S(t) = 1 ; K = k\right) & - \mathbb{P}\left(S(t-1) = 1 ; K = k\right) \\
 & = p(k)\left(\mathbb{P}\left(S(t) = 1 \big| K = k\right) -  \mathbb{P}\left(S(t-1) = 1 \big| K = k\right) \right). 
 \end{split}
  \end{align}
%
can also be interpreted as fraction of nodes with degree $k$ and failure time $t$ with respect to all nodes in the network. 

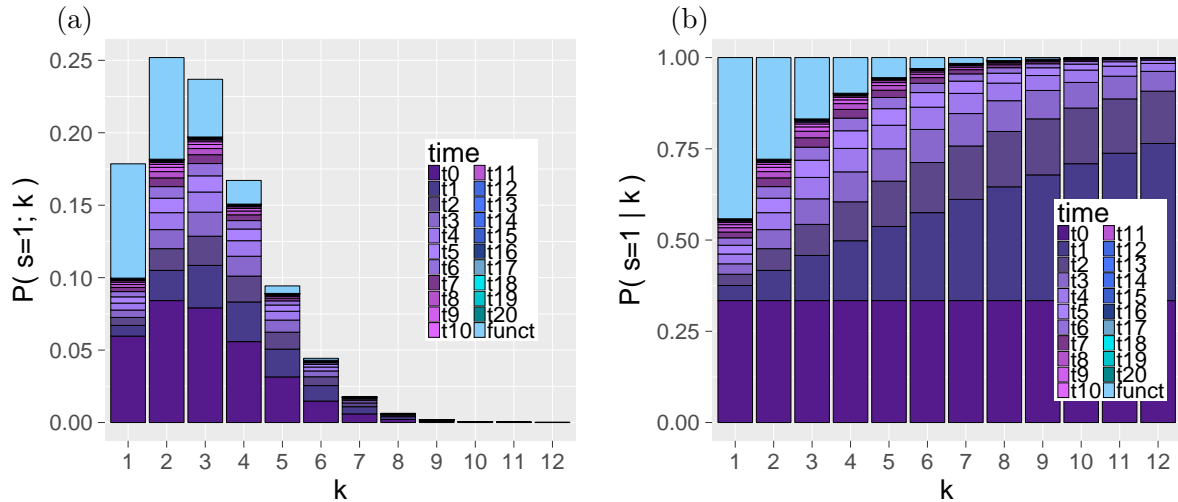
\begin{figure}[t]
 \centering
\includegraphics[width=0.49\textwidth]{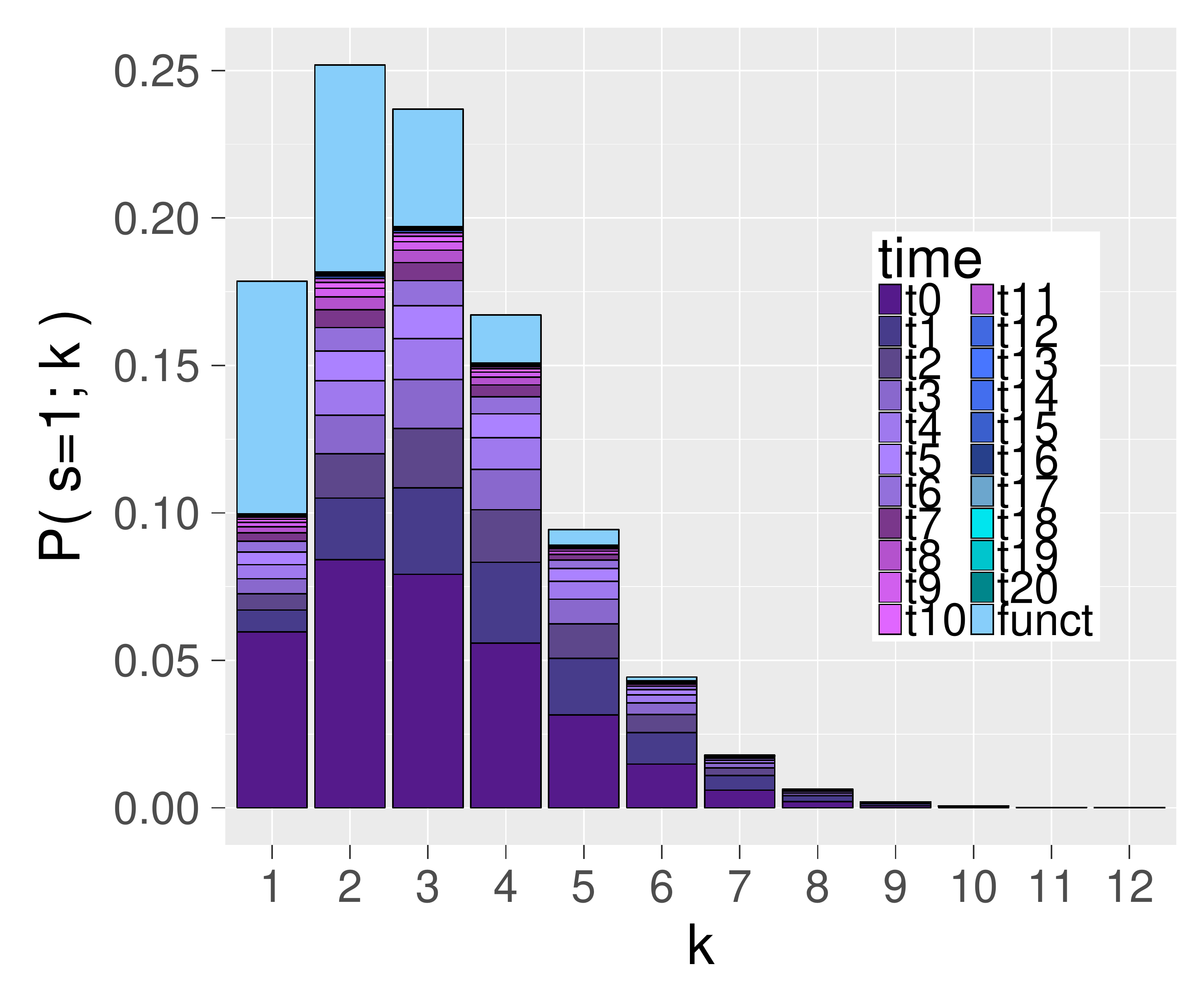}
\includegraphics[width=0.49\textwidth]{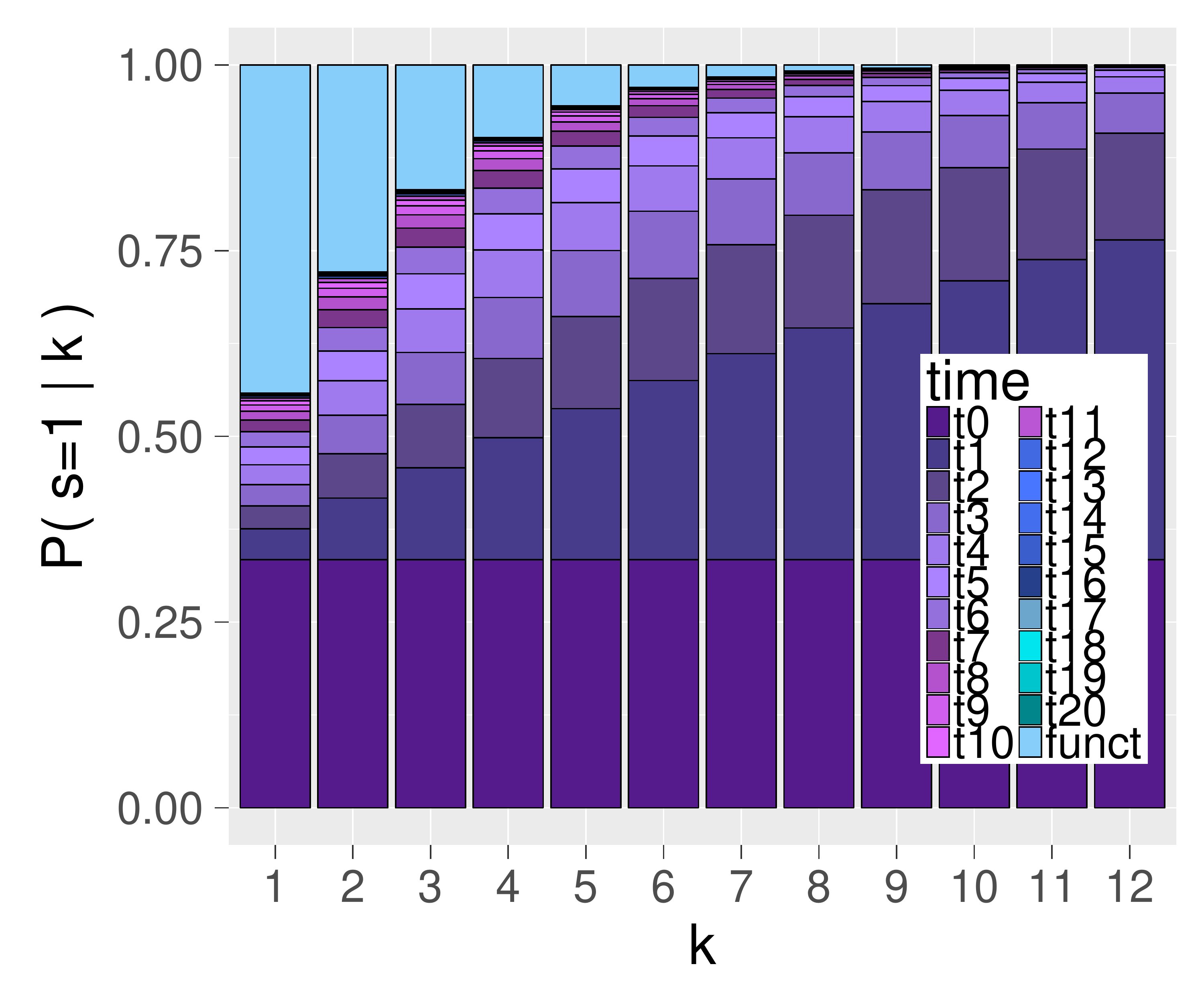}
\begin{picture}(0,0)
    \put(-430,185){(a)}
    \put(-200,185){(b)}
   \end{picture}
\caption{Time evolution of failures for Poisson random graphs (with average degree $z=3$) obtained by an LTA for the fiber bundle model with initial load $\lambda_0 = 0.5$. The thresholds are independently normally distributed with mean $\mu = 0.3 + \lambda_0$ and standard deviation $\sigma = 0.7$. The color indicates the time of failure of the respective fraction of failed nodes.
(a) Probability that a node has degree $k$ and is failed at the time indicated by the color.
(b) Probability that a node drawn uniformly at random from all nodes with degree $k$ fails at the time indicated by the color.
The light blue color corresponds to nodes that remain functional till the end of a cascade.
}
\label{fig:LRD_pfail}
\end{figure}

Both probabilities are illustrated in Fig.~\ref{fig:LRD_pfail} for the studied fiber bundle model and specific parameters.
Early on, the failure probability grows faster for nodes with a higher degree (Fig.~\ref{fig:LRD_pfail}~(b)).
Initially, all nodes fail with the same probability, but already at $t=1$, when the load distribution starts, high degree nodes are stronger impacted by a cascade.
Clearly, nodes with a higher degree have a higher failure risk, since more neighbors can possible fail and distribute load to them.
Yet, they tend to fail early and (almost) completely.
At later times, especially nodes with a smaller degree tend to fail.
Since these make up for the majority of nodes in the network, especially their failure would need to be prevented.  

\section{Conclusions}

Our main contribution is of methodological nature. 
We have presented a local tree approximation for specific fiber bundle models \citep{moreno2002instability,Lorenz2009,Tessone2012} on random graphs ensembles with prescribed degree distribution and degree-degree correlations.
Additionally, we have extended general constant load cascade models on weighted network ensembles \citep{Burkholz2015} to capture degree-degree correlations. 
Our derivations are exact in the thermodynamic limit of infinite network size, but approximate also large finite systems well.

Furthermore, we have shed light on the fact that we can capture the full cascade size dynamics and not only the final cascade outcome.
This is of great importance when interventions or other influencing factors in time shall be studied.
Or early warning signals might be found in the growth behavior or history of the average cascade.
However, it is worth noting that the dynamics span only a few time steps in our examples.

The analytic time resolution allows additionally to analyze the failure probability of nodes at specific times with respect to their degree.
The role that hubs or leaves play in the amplification of cascades can inspire further strategies to prevent or enhance the growth of cascades.

In this work, we have mainly provided a proof of concept that our derivations are exact. 
For the presented fiber bundle and constant load models, our framework enables directly the further study of degree-degree correlations, other degree and threshold distributions, correlations between such a threshold and degree distribution and other forms of intervention.

In fact, our framework applies to a much broader category of cascade models.
In comparison with known branching process approximations, we have introduced a perspective shift towards the iterative update of suitable probability distributions.
These distributions belong to random variables that describe the impact of nodes on their neighborhood that have failed at any point in the past and respect the Markovian nature of the studied random processes.
Several models that involve accumulation processes become analytically tractable this way.

We see potential to inspire the analytic investigation of even further classes of processes on networks.
%

%

%
%
%

%

\paragraph*{Acknowledgements.}
RB acknowledges support by the ETH48 project. RB and FS acknowledge financial support from the Project CR12I1\_127000
{\it OTC Derivatives and Systemic Risk in Financial Networks} financed by the Swiss National Science Foundation. FS acknowledges support by the EU-FET project MULTIPLEX 317532.

\bibliographystyle{sg-bibstyle}
\bibliography{Thesis}

\end{document}